\documentclass{lmcs}
\pdfoutput=1

\usepackage{lastpage}
\lmcsdoi{22}{3}{8}
\lmcsheading{}{\pageref{LastPage}}{}{}%
{Dec.~16,~2025}{Sep.~02,~2026}{}

\numberwithin{equation}{section}
\numberwithin{figure}{section}
\usepackage[utf8]{inputenc}

\usepackage{amssymb}
\usepackage{mathtools}
\usepackage{stmaryrd}
\usepackage{xspace}
\usepackage{hyperref}
\usepackage[nameinlink]{cleveref}

\usepackage[shortcuts]{extdash}

\usepackage{tikz}

\newcommand{\ident}[1]{\ensuremath{\texorpdfstring{\mathrm{\mathnm{#1}}}{#1}}\xspace}

\newcommand{\domain}[1]{\ensuremath{\mathbb{#1}}\xspace}

\newcommand{\N}{\domain{N}}

\newcommand{\eqdef}{\stackrel{\text{def}}=}

\renewcommand{\mod}[1]{\allowbreak\mkern10mu({\operator@font mod}\,\,#1)}

\newcommand{\mathcalsym}[1]{\ensuremath{\mathcal{#1}}\xspace}

\newcommand{\CA}{\mathcalsym{A}}

\newcommand{\CG}{\mathcalsym{G}}
\newcommand{\CH}{\mathcalsym{H}}

\newcommand{\from}{\colon}

\newcommand{\dom}{\ident{dom}}

\newcommand{\lang}{\ident{L}}

\usetikzlibrary{positioning}
\usetikzlibrary{decorations.pathreplacing}
\usetikzlibrary{automata,positioning}
\usetikzlibrary{decorations.pathmorphing}
\usetikzlibrary{decorations.markings}
\usetikzlibrary{decorations}
\usetikzlibrary{arrows}
\usetikzlibrary{patterns}
\usetikzlibrary{calc}
\usetikzlibrary{shapes}
\usetikzlibrary{fadings,	shadings}

\usetikzlibrary{arrows.meta}

	\tikzstyle{ubrace} = [draw, thick, decoration={brace, amplitude=7pt, mirror, raise=0.0cm}, decorate,
	    every node/.style={anchor=north, yshift=-0.15cm}]
	\tikzstyle{rbrace} = [draw, thick, decoration={brace, amplitude=7pt, mirror, raise=0.0cm}, decorate,
	    every node/.style={anchor=west, xshift= 0.15cm}]

	\tikzstyle{obrace} = [draw, thick, decoration={brace, amplitude=7pt, raise=0.0cm}, decorate,
	    every node/.style={anchor=south, yshift= 0.15cm}]
	\tikzstyle{lbrace} = [draw, thick, decoration={brace, amplitude=7pt, raise=0.0cm}, decorate,
	    every node/.style={anchor=east, xshift=-0.15cm}]

\newcommand{\eval}[2]{%
	\pgfmathparse{#2}%
	{\global\edef#1{\pgfmathresult}}%
}

\newcommand{\evalInt}[2]{%
	\pgfmathparse{int(#2)}%
	{\global\edef#1{\pgfmathresult}}%
}

\newcommand{\dL}{\mathtt{{\scriptstyle L}}}
\newcommand{\dR}{\mathtt{{\scriptstyle R}}}

\newcommand{\mathnm}[1]{#1}

\newcommand{\restr}{{\upharpoonright}}

\tikzstyle{dot} = [draw,shape=circle,fill, minimum size=1mm, inner sep=0pt, outer sep=0pt]
\tikzstyle{edge}  = [draw, thick,->]

\newenvironment{proofof}[1]
	{\vspace{1ex}\noindent{\emph{Proof of #1}}\hspace{0.5em}}
    {\hfill\qed\vspace{1ex}}

\newcommand{\init}{\ident{I}}

\DeclareSymbolFont{yhlargesymbols}{OMX}{yhex}{m}{n}
\DeclareMathAccent{\yhwidehat}{\mathord}{yhlargesymbols}{"62}

\newcommand{\trees}{\ident{Tr}}
\newcommand{\hist}{\ident{hist}}
\newcommand{\val}{\ident{val}}

\newcommand{\WF}{\ident{WF}}

\newcommand{\der}{\ident{der}}

\newcommand{\unravel}{\ident{unravel}}

\newcommand{\sT}{\mathsf{T}}
\newcommand{\sR}{\mathsf{R}}

\newcommand{\rank}{\ident{rank}}
\newcommand{\rankX}[1]{\ident{rank}^{#1}}
\newcommand{\rankR}{\ident{rank}^\sR}
\newcommand{\rankT}{\ident{rank}^\sT}
\newcommand{\apr}{\ident{apr}}

\renewcommand\phi\varphi
\renewcommand\epsilon\varepsilon

\newcommand{\eve}{\ensuremath{\exists}\xspace}
\newcommand{\adam}{\ensuremath{\forall}\xspace}

\newcommand{\ignore}[1]{}

\newcommand*\circled[1]{\tikz[baseline=(char.base),scale=0.6]{
            \node[shape=circle,draw,inner sep=0mm,scale=0.83,minimum size=3.7mm] (char) {\textbf{#1}};}}

\newcommand{\fb}{\mathbf{\text{\circled{2}}}}
\newcommand{\fg}{\mathbf{\text{\circled{1}}}}

\DeclarePairedDelimiter{\set}{\{}{\}}
\DeclarePairedDelimiter{\pair}{\llparenthesis}{\rrparenthesis}

\newcommand{\imwin}[2]{[#1\to #2]}

\Crefname{lem}{Lemma}{Lemmata}
\Crefname{cor}{Corollary}{Corollaries}
\Crefname{thm}{Theorem}{Theorems}
\Crefname{fact}{Fact}{Facts}
\Crefname{clm}{Claim}{Claims}
\Crefname{rem}{Remark}{Remarks}
\Crefname{exa}{Example}{Examples}
\Crefname{qu}{Question}{Questions}
\Crefname{prop}{Proposition}{Propositions}
\Crefname{equation}{Formula}{Formulae}
\Crefname{property}{Property}{Properties}\creflabelformat{property}{(#2#1#3)}

\newcommand{\dist}{\mathit{dist}}
\newcommand{\stab}{\mathit{st}}

\newcommand{\wit}{\ident{wit}_\omega}

\newcommand{\limit}{\ident{rank}}

\newcommand{\derF}{\der_\omega}

\newcommand{\limitR}{\mathrm{rank}}
\newcommand{\limitF}{\mathrm{rank}^{\sR}}

\tikzstyle{very loosely dotted}=[dash pattern=on \pgflinewidth off 6pt]

\newcommand{\gSta}[2]{
	\node[anchor=mid] at #1 {#2};
}

\newcommand{\gStT}[4]{
	\gSta{(#1*\ssx,#2*\ssy+#3*0.35)}{#4}
}

\newcommand{\gStR}[4]{
	\gSta{(#1*\ssx+\sss,#2*\ssy+#3*0.35)}{#4}
}

\newcommand{\gDot}[2]{
	\node[dot] (#1) at #2 {};
}

\newcommand{\gCon}[2]{
	\draw (#1) -- (#2);
}

\newcommand{\gConP}[2]{
	\draw (#1) edge[dashed] (#2);
}

\newcommand{\gLab}[3]{
	\node[anchor=mid east] at (-\ssx, #3*\ssy+0.5*\ssy) {#1:};
	\node[anchor=mid west] at (-\ssx, #3*\ssy+0.5*\ssy) {#2};
}

\newcommand{\gTta}[4]{
	\gDot{#1-T}{($(#1)+(+#2*0.3*\ssx,+1*\ssy)$)}
	\gDot{#1-R}{($(#1)+(-#2*0.3*\ssx,+1*\ssy)$)}

	\node[scale=0.8, anchor=mid east] at ($(#1)+(-0.2,-0.2)$) {$\delta_{#4}$};
	\node[scale=0.8, anchor=mid] at ($(#1)+(0,+0.95*\ssy)$) {#3};
}

\newcommand{\gTtb}[4]{
	\gTta{#1}{#2}{#3}{#4}
	
	\draw (#1) edge[line width=0.6mm, line cap=round] (#1-T);
	\draw (#1) edge[] (#1-R);
}

\newcommand{\gTtl}[4]{
	\gTta{#1}{#2}{#3}{#4}
	
	\draw (#1) edge[](#1-T);
	\draw (#1) edge[very loosely dotted](#1-R);
}

\newcommand{\gTrl}[4]{
	\gTta{#1}{#2}{#3}{#4}
	
	\draw (#1) edge[](#1-T);
	\draw (#1) edge[very loosely dotted](#1-R);
}

\newcommand{\itemfont}[1]{\textbf{#1}}


\begin{document}

\title[A~Dichotomy Theorem for Ordinal Ranks in MSO]{A~Dichotomy Theorem for Ordinal Ranks in MSO\rsuper*}
\titlecomment{{\lsuper*}This is a~journal version of a~paper that originally appeared at STACS 2025}

\author[D. Niwiński]{Damian Niwiński\lmcsorcid{0000-0002-1342-9805}}
\author[P. Parys]{Paweł Parys\lmcsorcid{0000-0001-7247-1408}}
\author[M. Skrzypczak]{Michał Skrzypczak\lmcsorcid{0000-0002-9647-4993}}

\address{Institute of Informatics, University of Warsaw, Poland}
\email{niwinski@mimuw.edu.pl, parys@mimuw.edu.pl, mskrzypczak@mimuw.edu.pl}

\thanks{All authors are supported by the National Science Centre, Poland (grant no.\@ 2021/\allowbreak41/\allowbreak B/\allowbreak ST6/\allowbreak03914).}

\begin{abstract}
	We focus on formulae $\exists X.\, \varphi(\vec{Y}, X)$ of monadic second\=/order logic over the full binary tree, such that the witness $X$ is a~well\=/founded set.
	The ordinal rank $\rank(X) < \omega_1$ of such a~set $X$ measures its depth and branching structure.
	We search for the least upper bound for these ranks, and discover the following dichotomy depending on the formula~$\varphi$.
	Let $\rank(\varphi)$ be the minimal ordinal such that, whenever an~instance $\vec{Y}$ satisfies the formula, there is a~witness $X$ with $\rank(X) \leq \rank(\varphi)$.
	Then $\rank(\varphi)$ is either strictly smaller than~$\omega^2$ or it reaches the maximal possible value~$\omega_1$.
	Moreover, it is decidable which of the cases holds.
	The result has potential for applications in a~variety of ordinal\=/related problems, in particular it entails a~result about the closure ordinal of a~fixed\=/point formula.
\end{abstract}

\maketitle

\section{Introduction}
\label{sec:intro}

The concept of a~well\=/founded relation plays a~central role in foundations of mathematics.
It gives rise to ordinal numbers, which underlie the basic results in set theory, for example that any two sets can be compared in cardinality.
Well\=/foundedness is no less important in the realm of computer science, where it often underlies the proofs of \emph{termination} of non\=/deterministic processes,
especially when no efficient bound on the length of a~computation is known.
In such cases, the complexity of possible executions is usually measured using an~ordinal called \emph{rank}.
Such a~rank can be seen as a~measure of the \emph{depth} of the considered partial order, taking into account suprema of lengths of possible descending chains.
Estimates on a~rank can provide upper\=/bounds on the computational complexity of the considered problem~\cite{schmitz_hab}.

In this work, we adopt the perspective of mathematical foundations of program verification and model\=/checking.
We focus on the monadic second\=/order logic (MSO) interpreted in the
infinite binary tree (with the left and right successors as the only
non\=/logical predicates), which is one of the reference formalisms in the area~\cite{thomas_languages}.
The famous Rabin Tree Theorem~\cite{rabin_s2s} established its decidability, but---half a~century after its introduction---the theory is still an~object of study.
On the one hand, it has led to numerous extensions, often shifting the decidability result far beyond the original theory (see e.g.~\cite{bojanczyk_wmso_u_p,ong_model_checking_higher_order}).
On the other hand, a~number of natural questions regarding Rabin's theory remain open, including a~large spectrum of \emph{simplification} problems.
For example, we still do not know whether we can decide if a~given formula admits an~equivalent form with all quantifiers restricted to finite sets.
Similar questions have been studied in related formalisms like $\mu$\=/calculus or automata;
for example, whether we can effectively minimise the Mostowski index of a~parity tree automaton~\cite{loding_index_to_bounds,murlak_game_auto},
or the $\mu \nu $\=/alternation depth of a $\mu$\=/calculus formula~\cite{bradfield_simplifying}.

On the positive side, some decidability questions have been solved by reductions to the original theory.
For example, it has been observed~\cite{niwinski_cardinality} that for a~given formula $\varphi(\vec{X})$,
the \emph{cardinality} of the family of tuples of sets $\vec{X}$ satisfying $\varphi(\vec{X})$ can be computed;
this cardinality can be either finite, $\aleph_0$, or $\mathfrak c$.
Later on, B\'ar\'any, Kaiser, and Rabinovich~\cite{barany_expressing_trees} proved a~more general result: they studied \emph{cardinality quantifiers} $\exists^{\geq \kappa} X.\, \varphi(\vec{Y}, X)$,
stating that there are at least $\kappa$ distinct sets $X$ satisfying $\varphi(\vec{Y}, X)$, and showed that these quantifiers can be expressed in the standard syntax of MSO;
thus the extended theory remains decidable.

In the present work, instead of asking \emph{how many} sets $X$ witness the formula $\exists X.\, \varphi(\vec{Y},X)$,
we ask how \emph{complex} these witnesses must be in terms of their depth\=/and\=/branching structure.
A~set $X$ of nodes of a~tree is \emph{well\=/founded} if it contains no infinite chain with respect to the descendant order.
In this case, a~countable ordinal number $\rank(X)$ is well defined (see \cref{sec:basic} below);
intuitively, the smaller $\rank(X)$, the simpler the set $X$ is, in terms of its \emph{branching structure}.

We consider formulae of the form $\exists X.\, \varphi(\vec{Y}, X)$, where $\varphi$ is an~arbitrary formula of MSO (it may contain quantifiers).
We assume that, whenever the formula is satisfied for some valuation of variables $\vec{Y}$, the value of $X$ witnessing the formula is a~well\=/founded set.
Note that well\=/foundedness of a~set is expressible in MSO (it suffices to say that each branch contains only finitely many nodes in $X$),
hence the requirement can be expressed within $\varphi$.
For a~fixed formula as above, we search for a~minimal ordinal $\limit(\phi)$ such that the rank of a~witness can be bounded by $\limit(\phi)$,
\begin{equation}\label{eq:eta-for-phi}
	\limit(\phi) \eqdef\adjustlimits\sup_{\vec{Y}}\min_{X}\, \rank(X),
\end{equation}
where $\vec{Y}$ and $X$ range, as expected, over the values satisfying $\varphi(\vec{Y}, X)$.

Since $\limit(\phi)$ is a~supremum of countable ordinals, its value is at most $\omega_1$ (the least uncountable ordinal).
This value can be achieved, for example, by the formula ``\emph{$X = Y$ and $Y$ is a~well\=/founded set}'',
as there are well\=/founded sets of arbitrarily large countable ranks.
On the other hand, for each pair of natural numbers $(k,l)$, one can construct a~formula $\varphi$ with $\limit(\phi) = \omega\cdot k+l$
in an~analogous way to Czarnecki~\cite{czarnecki_closure_fics}, see \cref{lem:construct-ordinals} below.
The main result of this work shows that no other ordinals can be obtained:

\begin{thm}\label{thm:main}
	For any formula $\exists X.\, \varphi(\vec{Y}, X)$ as above, the ordinal $\limit(\phi)$ is either strictly smaller than $\omega^2$ or equal to $\omega_1$.
	Moreover, it can be effectively decided which of the cases holds.
	In the former case, it is possible to compute a~number $N\in\N$ such that $\limit(\phi)< \omega\cdot N$.
\end{thm}
\noindent 
We also show that, in contrast to the aforementioned cardinality quantifiers, the property that $\rank(X)$ is smaller than $\omega^2$ cannot be expressed directly in MSO (see \cref{cor:not-expressible}).

The proof of \cref{thm:main} develops the game\=/based technique used previously to characterise certain properties of MSO\=/definable tree languages (see e.g.~\cite{clemente_separability,walukiewicz_buchi}).
Each application of this technique requires a~specific game, designed in a~way which reflects the studied property.
The game should have a~finite arena and be played between two perfectly\=/informed players \eve and \adam.
The winning condition of the game is given by a~certain $\omega$\=/regular set.
Then, the seminal result of Büchi and Landweber~\cite{buchi_synthesis} yields that the game is determined and the winner of this game can be effectively decided.
The construction of the game is such that a~winning strategy of each of the players provides a~witness of either of the considered possibilities; in our case:
if \eve wins then $\limit(\phi)=\omega_1$ and if \adam wins then $\limit(\phi) < \omega\cdot N$ for some computable $N$.
The heart of this approach lies in a~proper definition of the game, so that both these implications actually hold.

\subsection*{Related work}

In the context of $\mu$\=/calculus, one asks how many iterations are needed to reach a~fixed point;
this aspect concerns complexity of model checking (cf.~\cite{julian-igor-handook,esparza-acceleration}), as well as expressive power of the logic (cf.~\cite{julian-igor-handook,bradfield_transfinite}).
Recall that, in an~infinite structure, a~least fixed point $\mu X.\, F(X)$ is in general reached in a~transfinite number of iterations:
$\emptyset, F(\emptyset),\ldots, F^{\gamma}(\emptyset), \ldots$, where, for a~limit ordinal $\gamma$, $F^{\gamma}(\emptyset) = \bigcup_{\xi < \gamma} F^{\xi}(\emptyset)$.
It is therefore natural to ask if, for a~given formula, one can effectively find a \emph{closure ordinal} $\limit(F)$,
such that, in any model, the fixed point can be reached in $\limit(F)$, but, in general, not in fewer iterations.
This notion generalises in a~natural way to a~vectorial least fixed point $\mu \vec{X}.\,\vec{F}(\vec{X})$,
where $\vec{F}(\vec{X})$ is a~tuple of formulae $F_1(\vec{X}),\dots,F_k(\vec{X})$ and $\vec{X}$ is a~tuple of variables $X_1,\dots,X_k$.
Fontaine~\cite{fontaine_continuous} effectively characterised the formulae such that in each model the fixed point is reached in a~finite number of steps.
Czarnecki~\cite{czarnecki_closure_fics} observed that some formulae have no closure ordinals but, for each ordinal\footnote{Czarnecki considered only infinite ordinals, but the claim holds also for $\eta < \omega$. For example, a~formula $\mu X.\, \left(\Box^5 \bot \wedge(\Diamond X \vee \Box \bot) \right)$ has closure ordinal $5$.} $\eta < \omega^2$, there is a~formula whose closure ordinal is $\eta$.
He also raised the following question.

\begin{qu}[{Czarnecki~\cite{czarnecki_closure_fics}}]\label{con:czarnecki}
	Is there a $\mu$\=/calculus formula of the form $\mu X.\,F(X)$ that has a~countable closure ordinal $\limit(F)\geq\omega^2$?
\end{qu}

Gouveia and Santocanale~\cite{santocanale-closure} exhibited an~example of a~formula with an~essential alternation of the least and greatest fixed\=/point operators whose closure ordinal is $\omega_1$; clearly this limit can be achieved only in uncountable models.
In general, it remains open whether a~formula $\mu X.\,F(X)$ of the $\mu$\=/calculus may have a \emph{countable} closure ordinal $\limit(F) \geq \omega^2$.
Afshari and Leigh~\cite{afshari_closure} claimed a~negative answer for formulae of the alternation\=/free fragment of $\mu$\=/calculus;
however, as the authors have later admitted~\cite{afshari_limit_fics}, the proof contained some gaps.
In a~recent paper~\cite{afshari_limit_fics}, Afshari, Barlucchi, and Leigh update the proof and extend the result to formulae of the so\=/called \emph{$\Sigma$\=/fragment} of the $\mu$\=/calculus.
More specifically, the authors consider vectorial fixed points $\mu \vec{X}.\,\vec{F}(\vec{X})$,
where the formulae in $\vec{F}$ may contain closed sub\=/formulae of the full $\mu$\=/calculus, but the variables of $\vec{X}$ do not fall in the scope of any fixed\=/point operators.
The authors show that if a~countable number of iterations $\eta$ suffices to reach the least fixed point of such a~system in any countable Kripke frame then $\eta < \omega^2$.

As a~direct consequence of our result, we obtain an~alternative proof of the same theorem.
This is achieved via a~well\=/known reduction of the $\mu$\=/calculus to the MSO theory of the~binary tree,
based on the tree\=/model property~\cite{julian-igor-handook}, and a~natural encoding of a~tree model.
Compared with Afshari, Barlucchi, and Leigh, we additionally obtain that it can be decided whether $\limit(\vec{F}) < \omega^2$ or $\limit(\vec{F})\geq\omega_1$.
Another difference is that our definition of $\limit$ allows arbitrary models, while Afshari, Barlucchi, and Leigh consider only countable models;
however, we prove that this does not change anything: if a~formula requires some countable number of iterations in some (uncountable) model, then the same holds in some countable model.
Yet another difference is that the proof of Afshari, Barlucchi, and Leigh works only for so\=/called guarded formulae.
While it is known~\cite{kozen-mu-calc} that every formula can be converted to an~equivalent guarded one, this does not have to preserve the closure ordinal
(note that the closure ordinal is assigned to a~formula, not to the property expressed by the formula); we deal with this issue in \cref{lem:guarded}.

Similar dichotomies have been discovered in the studies of decision problems related to topological complexity of MSO\=/definable tree languages.
An~open problem related to the aforementioned question of definability with finite\=/set quantifiers, is whether we can decide if a~tree language belongs to the Borel hierarchy (in general, it need not).
A~positive answer is known if a~tree language is given by a~deterministic parity automaton~\cite{niwinski_gap}, based on the following dichotomy:
such a~language is either $\Pi^1_1$\=/complete (very hard), or on the level $\Pi^0_3$ (relatively low) of the Borel hierarchy.
Skrzypczak and Walukiewicz~\cite{walukiewicz_buchi} gave a~proof in the case when a~tree language is given by a~non\=/deterministic Büchi tree automaton,
inspired by a~rank\=/related dichotomy conjectured in the previous work of the first author~\cite[Conjecture~4.4]{mskrzypczak_lncs}.
There, an~ordinal has been associated with each Büchi tree automaton, and it turns out (in view of~\cite{walukiewicz_buchi}) that this ordinal either equals $\omega_1$ or is smaller than $\omega^2$,
in which case the tree language is Borel.
It should be mentioned that a~procedure to decide whether a~Büchi definable tree language is weakly definable (without the topological counterpart) was given earlier by Colcombet et al.~\cite{colcombet_weak}.

Another perspective on closure ordinals is proposed in Bruse, S\"{a}lzer, and Lange~\cite{bruse_closure}, where the authors consider a~fixed structure but varying formulae.
The main result states that there exists an~infinite word that is not ultimately periodic, yet every formula reaches its fixed point in a~finite number of steps. See also~\cite{lange-convergence} for further results in this direction.

The relation between ordinals and automata has also been considered in studies of automatic structures.
In particular, Delhomm\'e~\cite{delhomme} showed that automatic ordinals are smaller than $\omega^{\omega}$ while tree\=/automatic ordinals (defined in terms of automata on finite trees) can go higher,
but not above $\omega^{\omega^{\omega}}$.
Later, Finkel and Todorcevic~\cite{finkel-todo-auto} showed that $\omega$\=/tree automatic ordinals are smaller than $\omega^{\omega^{\omega}}$ as well.
These results may appear in contrast with our more restrictive bound of $\omega^2$;
however, representation by automatic structures is in~general more powerful than expressibility in MSO, so the two approaches are not directly related.

The present paper is an~extended version of a~conference paper~\cite{dichotomy-stacs}. Compared with the conference version, we base the proof of dichotomy for $\rank(\cdot)$ on another rank (denoted $\rankR(\cdot)$), for which we are able to additionally show that its values can be exactly computed (see \cref{pro:main-layers} below). Moreover, we develop tools to study closure ordinals for vectorial fixed points, so that our results on MSO can be applied to them. This implies the dichotomy for vectorial fixed points (see \cref{thm:closure-mu} below).

\section{Basic notions}\label{sec:basic}

$\N=\{0,1,2,\ldots\}$ denotes the set of natural numbers.
We use standard notation for ordinal numbers, with $0$ being the least ordinal number, $\omega$ being the least infinite ordinal, and $\omega_1$ the least uncountable ordinal.
Although $\omega$ and $\N$ coincide as sets, we distinguish the two notations to emphasise the perspective.

Recall that ordinals can be multiplied.
Intuitively, $\gamma\cdot\eta$ represents $\eta$ many copies of a~well\=/ordered set of type $\gamma$, arranged one after another.
For example, we have $\omega\cdot 2=\omega+\omega$ and $2\cdot\omega=\omega$.

\paragraph*{Words.}
An~alphabet $A$ is any finite non\=/empty set, whose elements are called \emph{letters}.
A~\emph{word} over $A$ is a~finite sequence of letters $w=w_0\cdots w_{n-1}$, with $w_i\in A$ for $i <n$, and $n$ being the \emph{length} of $w$.
The empty word, denoted $\epsilon$, is the unique word of length $0$.
By $u v$ we denote the \emph{concatenation} of words $u=u_0\cdots u_{n-1}$ and $v=v_0\cdots v_{m-1}$, that is, $u v = u_0\cdots u_{n-1} v_0\cdots v_{m-1}$.

$A^\ast$ denotes the set of all finite words over the alphabet $A$, while $A^\omega$ denotes the set of all $\omega$\=/words over $A$, that is, functions $\alpha\from \N\to A$.
The \emph{length} of such a~word is $\omega$.

If $\alpha$ is a (finite or infinite) word and $n\in\N$ is less than or equal to the length of $\alpha$ then by~$\alpha\restr_n$
we denote the finite word consisting of the first $n$ letters of $\alpha$, that is, $\alpha_0\cdots\alpha_{n-1}\in A^\ast$.

The prefix order on words is defined as follows: for $u\in A^\ast$ and $v\in A^\ast\cup A^\omega$, we say that $u$ is a~\emph{prefix} of $v$, denoted $u\preceq v$, if there exists $n\leq |v|$ such that $v\restr_n = u$.

\paragraph*{Trees.}
Let $\dL$ and $\dR$ denote two distinct letters called \emph{directions}.
For a~direction $d\in\{\dL,\dR\}$, the \emph{opposite direction} is denoted $\bar{d}\neq d$.
Words over the alphabet $\{\dL,\dR\}$ are called \emph{nodes}, with the empty word $\epsilon$ often called the \emph{root}.
An~$\omega$\=/word $\alpha\in\{\dL,\dR\}^\omega$ is called an~\emph{infinite branch}.
A~node $u$ is called the \emph{parent} of its \emph{children} $u \dL$ and $u\dR$.

A~(full infinite binary) \emph{tree} over an~alphabet $A$ is a~function $t\from \{\dL,\dR\}^\ast\to A$, assigning letters $t(u)\in A$ to nodes $u\in\{\dL,\dR\}^\ast$.
The set of all trees over $A$ is denoted $\trees_A$.
A~\emph{subtree} of a~tree $t\in\trees_A$ rooted at a~node $u\in\{\dL,\dR\}^\ast$ is the tree $t\restr_u$ over $A$ defined by taking $t\restr_u(v) = t(uv)$.

\paragraph*{Automata.}
Instead of working with formulae of MSO, we use another equivalent formalism, namely non\=/deterministic parity tree automata~\cite{thomas_languages}.
An~\emph{automaton} is a~tuple $\CA=(A,Q, q_{\init}, \Delta,\Omega)$, where $A$ is a~finite alphabet, $Q$ is a~finite set of \emph{states}, $q_{\init}\in Q$ is an~\emph{initial state},
$\Delta\subseteq Q\times A\times Q\times Q$ is a~\emph{transition relation}, and $\Omega\from Q\to \{i,\ldots,j\}\subseteq \N$ is a~\emph{priority mapping}.
A~\emph{run} of an~automaton~$\CA$ from a~state $q\in Q$ over a~tree $t\in\trees_A$ is a~tree $\rho\in\trees_Q$ such that $\rho(\epsilon)=q$ and for every node $u\in\{\dL,\dR\}^\ast$ the quadruple
\[\big(\rho(u), t(u), \rho(u\dL),\rho(u \dR)\big),\]
is a~transition of $\CA$ (i.e.,~belongs to $\Delta$).

A~sequence of states $(q_0,q_1,\ldots)\in Q^\omega$ is \emph{accepting} if $\limsup_{n\to\infty} \Omega(q_n)$ is an~even natural number.
A~run $\rho$ is \emph{accepting} if for every infinite branch $\alpha\in\{\dL,\dR\}^\omega$ the sequence of states $\big(\rho(\alpha\restr_n)\big)_{n\in\N}$ that appear in $\rho$ on $\alpha$ is accepting.

A~tree $t\in\trees_A$ is \emph{accepted} by an~automaton $\CA$ from a~state $q\in Q$ if there exists an~accepting run $\rho$ of $\CA$ from the state $q$ over~$t$.
The \emph{language} of $\CA$, denoted $\lang(\CA)\subseteq\trees_A$, is the set of trees which are accepted by the automaton from the initial state $q_\init$.
A~language $L\subseteq\trees_A$ is \emph{regular} if it is $\lang(\CA)$ for some automaton $\CA$.

We now recall the famous theorem of Rabin, which allows us to transform MSO formulae into equivalent tree automata.

\begin{thmC}[\cite{rabin_s2s}, see also~\cite{thomas_languages}]
	For every MSO formula $\phi(X_0,\ldots,X_{n-1})$ the set of valuations satisfying $\phi$ is a~regular language over the alphabet $\{0,1\}^n$.
\end{thmC}
\noindent 
We say that an~automaton is \emph{pruned} if every state $q\in Q$ occurs in some accepting run from the initial state, that is,
there exists an~accepting run $\rho$ of $\CA$ from $q_{\init}$ over a~tree $t$ such that $\rho(u)=q$ for some node $u\in\{\dL,\dR\}^\ast$.
Note that every automaton can effectively be pruned, without affecting its language, by detecting and removing states that do not appear in any accepting run.

\paragraph*{Games.}
We use the standard framework of two\=/player games of infinite duration (see e.g.,~\cite{buchi_synthesis,fijalkow_games}).
The arena of such a~game is given as a~graph with both players having perfect information about the current position.
The winning condition of the game is given by a~language of infinite plays won by one of the players.

\paragraph*{Ordinal ranks.}
The set $\{\dL,\dR\}^\ast$,
equipped with the prefix order, is obviously well\=/founded: it does not admit any infinite strictly descending chain $v_0 \succ v_1 \succ v_2 \succ \ldots$.
When discussing well\=/foundedness in the MSO theory of the full binary tree, it is therefore more fruitful to consider the reverse ordering
$\succeq$, which also coincides with the usual intuition of the tree growing downward.
Then a~set $X \subseteq \{\dL,\dR\}^\ast$ is well\=/founded precisely when all of its linearly ordered subsets are finite
(i.e., there are also no infinite strictly ascending chains $v_0 \prec v_1 \prec v_2 \prec \ldots$).
As we adopt the automata\=/based perspective, we do not work directly with tuples of
sets of nodes but rather with labellings of the tree over suitable finite alphabets. In particular, any set of nodes
$X\subseteq \{\dL,\dR\}^\ast$ can be identified with its characteristic function, viewed as the~tree $x\in\trees_{\{0,1\}}$ over the alphabet $\{0,1\}$.
We then apply the terminology for sets directly to such labelled trees.
The \emph{empty tree} is the unique tree $x_0$, such that $x_0(u)=0$, for all $u\in\{\dL,\dR\}^\ast$.
A~tree is \emph{finite} if $x(u)=1$ for only finitely many $u$.
A~tree $x\in\trees_{\{0,1\}}$ is \emph{well\=/founded} if the inverse image of $1$ is well\=/founded, that is, no branch contains infinitely many nodes labelled by $1$.
We denote by $\WF\subseteq\trees_{\{0,1\}}$ the
set of all well\=/founded trees.

For any tree $x \in \WF \subseteq \trees_{\{0,1\}}$ we define its \emph{rank} as an~ordinal
\[ 0 \leq \rank(x) < \omega_1. \]
There are several equivalent ways to introduce this notion; for the sake of later generalisations, we use a~definition based on the concept of a~\emph{derivative}.

We call a~function $d \colon \trees_{\{0,1\}} \to \trees_{\{0,1\}}$ a~\emph{derivative} if for every $x \in \trees_{\{0,1\}}$ and every $u \in \{\dL,\dR\}^\ast$ we have
\[ d(x)(u) \leq x(u). \]
We then consider the transfinite iteration of $d$, defined by
\[
d^0(x) = x,\qquad
d^{\gamma+1}(x) = d\bigl(d^\gamma(x)\bigr),
\]
and for limit ordinals $\eta$,
\[
d^\eta(x)(u) = \lim_{\gamma \to \eta} d^\gamma(x)(u),
\]
where the limit exists by monotonicity of $d$.

We are interested in derivatives for which this iteration reaches the empty tree at some stage; such a~stage is necessarily countable.

Now we apply this construction to the derivative $\der$ defined by
\[
\der(x)(u) \;\eqdef\;
\begin{cases}
  0	& \text{if $x\!\restriction_u$ contains at most one occurrence of $1$,} \\
  x(u)	& \text{otherwise,}
\end{cases}
\]
and we set
\[
\rank(x) \;\eqdef\; \inf\bigl\{\eta < \omega_1 \mid \der^\eta(x) \text{ is empty}\bigr\}.
\]
\noindent 
It is a~standard result that $\rank(x)$ is well defined if and only if $x \in \WF$ \cite{kechris_descriptive}.

An~alternative definition can be given in terms of counting functions.
If a~tree $x\in\trees_{\{0,1\}}$ is a~characteristic function of a~set $X \subseteq \{\dL,\dR\}^\ast$ that is well\=/founded with respect to~the relation $\succeq$,
then $\rank(x)$ is the least ordinal for which there exists a~monotone function $f\colon X \to \rank(x)$ that preserves the strict ordering, that is, $w \succ v$ implies $f(w) < f(v)$
(recall that, by definition, an~ordinal is a~set of all strictly smaller ordinals).
This is the definition we have used in the conference version of the paper~\cite{dichotomy-stacs}.

\begin{exa} \label{przyklad-grzebien}
	The tree $x_0\in\trees_{\{0,1\}}$ with all nodes labelled by $0$ has rank $0$.

	Consider a~tree $x_\omega$ where a~node $v$ has label $1$ if $v=\dR^i\dL^j$ with $1\leq j\leq i$.
	It is a~\emph{comb}: the rightmost branch is labelled by zeros, and below its $i$\=/th node we have a~tooth of $i$ nodes labelled by ones, going left.
	The rank of $x_\omega$ is $\omega$.

	Such combs can be nested: suppose that $x_{\omega\cdot 2}$ starts analogously to $x_\omega$, but below every tooth we again insert $x_\omega$ (i.e.,~$x_{\omega\cdot 2}\restr_{\dR^i\dL^{i+1}}=x_\omega$ for every $i$);
	then $\rank(x_{\omega\cdot 2})=\omega\cdot 2$.
	Repeating this, we can insert $x_{\omega\cdot n}$ below every tooth of $x_\omega$, and obtain $x_{\omega\cdot(n{+}1)}$ of rank $\omega\cdot(n{+}1)$, for every $n\in\N$.
	
	Then, we can place every $x_{\omega\cdot n}$ at node $\dR^n\dL$, below a~$0$\=/labelled rightmost branch; call the resulting tree $x_{\mathit{diag}}$.
	Then $\rank(x_{\mathit{diag}}) = \omega^2$.
	In a~similar manner we can create a~tree having rank equal to any countable ordinal $\eta$.
\end{exa}

\paragraph{Variants of ranks.}
The derivative considered above removes each occurrence of $1$ along a~path individually.
As a~result, $\rank(x)$ counts, in a~sense, all labels $1$ in the tree $x$.
For our purpose, it is useful to work with a~``faster'' derivative that can remove more $1$'s at once, provided that none of them has infinitely many descendants labelled $1$.
Define
\begin{align*}
	\derF(x)(u) &\eqdef
	\begin{cases}
	  0	& \text{if $x\restr_u$ contains finitely many $1$'s,} \\
	  x(u)	& \text{otherwise.}
	\end{cases}
\end{align*}
For this derivative, we define an~analogue of the previous rank:
\begin{align*}
	\rankR(x)&\eqdef\inf \{\eta<\omega_1\mid \text{$\derF^\eta(x)$ is empty}\}.
\end{align*}
To distinguish it from the original rank, we call this rank the \emph{layered depth} of $x$.

To see that $\rankR(x)$ is well defined for every $x' \in \WF$, observe that $\derF(x')(u) \leq \der(x')(u)$ for all $u\in\{\dL,\dR\}^\ast$. Consequently, $\rankR(x) \leq \rank(x)$.
For later use, we also introduce a~closely related rank,
\begin{align*}
	\rankT(x)&\eqdef\inf \{\eta<\omega_1\mid \text{$\derF^\eta(x)$ is finite}\}.
\end{align*}
Again, it is well defined for every $x \in \WF$ as, by definition, $\rankT(x) \leq \rankR(x)$.
The letters $\sR$ and $\sT$ derive from the words \emph{reach} and \emph{trunk}, which refer to two key mechanisms in the proof of our main result.
The intuition behind these two ranks becomes clear later.

\begin{exa}
	Let us reconsider the trees of \cref{przyklad-grzebien}.
	It is not hard to see that $\rankR(x_\omega) = \rankT(x_\omega) = 1$, $\rankR(x_{\omega\cdot 2})=\rankT(x_{\omega \cdot 2}) = 2$
	and, generally, $\rankR(x_{\omega\cdot n})=\rankT(x_{\omega \cdot n}) = n$.
	Consequently,
	\[
	\rankR(x_{\mathit{diag}}) =\rankT(x_{\mathit{diag}}) = \omega.
	\]
	Now consider a~tree $x^1_\omega$ that differs from $x_\omega$ only in that it has 1 in the root.
	It is not hard to see that $\rank(x^1_\omega)=\omega + 1$, and the new ranks are
	\[
	  \rankR(x^1_\omega)=2, \qquad \rankT(x^1_\omega)=1.
	\]
	To see that all three ranks may coincide, we can take, of course, the empty tree $x_0$.
	But it is also the case for the \emph{flat comb} tree $x_f$, where a~node $v$ has label $1$ if $v=\dR^i\dL$;
	we have $\rank(x_f) = \rankR(x_f) = \rankT(x_f) = 1$.
	By nesting this tree like in \cref{przyklad-grzebien}, we can eventually produce a~tree having all three ranks equal to any countable ordinal $\eta<\omega_1$.
\end{exa}

The following fact summarises the relationship between various ranks.

\begin{fact}\label{ft:monotone}
	For every well\=/founded tree $x\in\trees_{\{0,1\}}$ we have
	\begin{alignat}{4}
	\rankR(x) &\leq \rank(x) &&\leq\omega\cdot \rankR(x),\label{eq:r-leq-f-omega}\\
	\rankT(x) &\leq \rankR(x) &&\leq\rankT(x)+1.\label{eq:o-leq-f-leq-op}
	\end{alignat}
\end{fact}

\begin{proof}
	The first part of Inequality~\eqref{eq:r-leq-f-omega} was already noted when we introduced $\rankR$.
	The second part of Inequality~\eqref{eq:r-leq-f-omega} follows from the fact that $\der^\omega(x')(u) \leq \derF(x')(u)$, for every $x'\in\WF$ and $u\in\{\dL,\dR\}^\ast$.
	Inequalities~\eqref{eq:o-leq-f-leq-op} come directly from the definitions.
\end{proof}

We use the following two lemmata connecting the values $\rankR(x)$ and $\rankT(x)$.

\begin{lem}\label{lem:inf-fank-to-oank}
	Let $\gamma<\omega_1$ be a~countable ordinal number.
	Take $x\in\WF$ and consider infinitely many nodes $w_0,w_1,\ldots\in\{\dL,\dR\}^\ast$ which form an~anti\=/chain (for $i\neq j$ we have neither $w_i\preceq w_j$ nor $w_j\preceq w_i$).
	If for every $i\in\N$ we have $\rankR(x\restr_{w_i})\geq \gamma$ then $\rankT(x)\geq\gamma$.
\end{lem}

\begin{proof}
	Assume for the sake of contradiction that $\eta=\rankT(x)<\gamma$.
	Consider the tree $\derF^\eta(x)$.
	For every $i\in\N$ we have $\rankR(x\restr_{w_i})\geq\gamma>\eta$, which means that the tree $\derF^\eta(x)\restr_{w_i}$ is not empty.
	Therefore, as the $w_i$'s form an~infinite antichain, the tree $\derF^\eta(x)$ is not finite.
	Therefore, $\rankT(x)>\eta$; a~contradiction.
\end{proof}

\begin{lem}\label{lem:fin-oank-to-fank}
	Take $x\in\WF$ and consider a~pair of nodes $w\prec u\in\{\dL,\dR\}^\ast$.
	If $x(w)=1$ then $\rankR(x)\geq \rankT(x\restr_u)+1$.
\end{lem}

\begin{proof}
	Let $\gamma$ be the minimal ordinal such that $\derF^\gamma(x)(w)=0$.
	Clearly $\rankR(x)\geq\gamma$.
	Due to the definition of the iterations of a~derivative, we know that $\gamma$ is a~successor ordinal.
	Thus, $\derF^{\gamma-1}(x)(w)=1$ and $\derF^{\gamma-1}(x\restr_u)$ is finite (because another application of $\derF$ makes $1$ at $w$ disappear).
	This means that $\gamma-1\geq \rankT(x\restr_u)$ and therefore $\gamma\geq\rankT(x\restr_u)+1$.
\end{proof}

\section{Problem formulation}
\label{sec:problem}

We now re\=/state the problem under consideration in the automata\=/theoretic framework.
Rather than working with a~formula $\phi(\vec{Y},X)$, we consider a~regular tree language $\Gamma$ over an~alphabet $A\times\{0,1\}$, for a~suitable finite alphabet $A$.
We identify a~tree $\tau$ over $A\times\{0,1\}$ with a~pair of trees $(t,x)$, where $t\in\trees_A$, $x\in\trees_{\{0,1\}}$, and $\tau(u)=\big(t(u),x(u)\big)$, for every node $u\in\{\dL,\dR\}^\ast$.
Thus, $\Gamma\subseteq\trees_{A\times\{0,1\}}$ can be seen as a~relation whose elements are pairs $(t,x)$.
We additionally require that whenever $(t,x)\in\Gamma$ then $x$ is well\=/founded, which means that $\Gamma$ (treated as a~relation) is contained in $\trees_A\times \WF$.
We say that such a~relation is \emph{regular} if it is regular as a~tree language over $A\times\{0,1\}$.

Let $\pi_A(\Gamma)$ be the projection of $\Gamma$ onto the $A$ coordinate, that is,
the set of those trees $t\in\trees_A$, for which there exists a~(necessarily well\=/founded) tree $x\in\trees_{\{0,1\}}$, such that $(t,x)\in \Gamma$.
Similarly, for a~tree $t\in\trees_A$, by $\Gamma_t$ we denote the \emph{section} of $\Gamma$ over the tree $t$, that is, the set $\big\{x\in\trees_{\{0,1\}}\mid (t,x)\in\Gamma\big\}$.

The following definition is just a~reformulation of \cref{eq:eta-for-phi} in terms of a~relation $\Gamma$.

\begin{defi}
	The \emph{closure ordinal} and the \emph{layered ordinal} of a~relation $\Gamma\subseteq \trees_A{\times} \WF\subseteq \trees_{A\times\{0,1\}}$ are
	\begin{align*}
	\limitR(\Gamma) &\eqdef\adjustlimits\sup_{\ t\in\pi_A(\Gamma)\ }\min_{x\in\Gamma_t}\, \rank(x),\\
	\limitF(\Gamma) &\eqdef\adjustlimits\sup_{\ t\in\pi_A(\Gamma)\ }\min_{x\in\Gamma_t}\, \rankR(x).\\
	\end{align*}
	Both these values are called \emph{ordinals} of $\Gamma$.
	When referring to an~ordinal of an~automaton $\CA$, we mean the respective ordinal of $L({\CA})$.
\end{defi}

Due to the estimates from \cref{ft:monotone}
we know that $\limitF(\Gamma)\leq \limitR(\Gamma) \leq \omega\cdot \limitF(\Gamma)$.
Thus, to prove \cref{thm:main}, it is enough to show the following.

\begin{prop}\label{pro:main-layers}
	If $\Gamma\subseteq \trees_A{\times} \WF\subseteq \trees_{A\times\{0,1\}}$ is MSO definable then either $\limitF(\Gamma)< \omega$ or $\limitF(\Gamma)=\omega_1$.
	Moreover, it is decidable which of the cases holds and if $\limitF(\Gamma)<\omega$ then its concrete value can be effectively computed.
\end{prop}

\noindent 
The crucial advantage of \cref{pro:main-layers} over \cref{thm:main} comes from the fact that we work with $\limitF(\Gamma)$ here.
The provided construction allows us to compute precisely the layered ordinal $\limitF(\Gamma)$, while the values of $\limit(\Gamma)$ can only be estimated as in Inequality~\eqref{eq:r-leq-f-omega}.

\begin{exa}\label{ex:automaton}
	Consider the following automaton $\CA$ over the alphabet $\{\mathsf{b},\mathsf{c}\}\times\{0,1\}$.
	Its states are $p_0$, $q_1$, $q_2$, $q_3$, $r_0$, $r_1$ with $q_1$ being initial, where the subscript provides the priority of a~state (i.e.,~$\Omega(p_i)=\Omega(q_i)=\Omega(r_i)=i$).
	The transitions are, for all $i\in\{0,1\}$, $j\in\{1,2,3\}$, and $a\in\{\mathsf{b},\mathsf{c}\}$,
	\begin{align*}
		&(p_0,(a,0),p_0,p_0),&		&(q_j,(\mathsf{c},0),q_3,p_0),&	&(q_j,(\mathsf{c},0),p_0,r_1),\\
		&(q_j,(\mathsf{b},0),q_2,p_0),&	&(q_j,(\mathsf{c},0),p_0,q_3),&	&(r_i,(\mathsf{b},i),r_1,r_0),\\
		&(q_j,(\mathsf{b},0),p_0,q_1),&	&(q_j,(\mathsf{c},0),r_1,p_0),&	&(r_i,(\mathsf{c},0),p_0,p_0).
	\end{align*}
	In this example, we should see $\mathsf{c}$\=/labelled nodes as separators, splitting the whole tree into $\mathsf{b}$\=/labelled zones.
	Given a~tree $t\in\trees_A$, let us see, for which witnesses $x$, if any, the pair $(t,x)\in\trees_{\{\mathsf{b},\mathsf{c}\}\times\{0,1\}}$ can be accepted.

	Note first that $\CA$ accepts $(t,x_0)$ from $p_0$ for every tree $t$ and for $x_0$ having all nodes labelled by $0$.
	Also note that states $q_j$ become aligned along a~single branch, which can be either finite or infinite.
	We can further see that $(t,x_0)$ can be accepted from the initial state $q_1$ if there exists a~branch $\alpha$ in $t$ satisfying the following property:
	$\alpha$ goes left infinitely often, but visits only finitely many $\mathsf{c}$\=/labelled nodes.
	Indeed, if we align states $q_j$ precisely with the branch $\alpha$ then the state $q_3$ is visited only finitely often (below every $\mathsf{c}$),
	and the state $q_2$ infinitely often, while turning left below $\mathsf{b}$, hence the parity condition is satisfied
	(all the remaining nodes hold the state $p_0$).
	Conversely, it is easy to see that whenever the states $q_i$ are aligned along an~infinite branch $\alpha$ (in an~accepting run) then this branch satisfies the aforementioned property.

	Another possibility is that the branch with states $q_j$ is finite, and below some $\mathsf{c}$\=/labelled node the state changes to $r_1$.
	Now, in $\mathsf{b}$\=/labelled nodes, the run sends $r_1$ to every left child, and $r_0$ to every right child;
	hence every left child in $x$ (and the root of the zone) should have label $1$, and every right child---label $0$.
	We can continue the zone of states $r_i$ until reaching a~node labelled by $\mathsf{c}$ in $t$;
	such a~node allows us to change the state into $p_0$ and accept anything below.
	The acceptance condition requires that there are only finitely many states $r_1$ (hence also nodes with label $1$ in $x$) on every branch;
	the tree $x$ is necessarily well\=/founded.
	
	This determines the optimal rank of a~witness~$x$ for a~tree~$t$.
	Namely, if in $t$ there is a~branch that infinitely often goes left, but visits only finitely many $\mathsf{c}$\=/labelled nodes, we have a~witness of rank $0$.
	Otherwise, we should consider every zone of $\mathsf{b}$\=/labelled nodes in~$t$, surrounded by $\mathsf{c}$\=/labelled nodes;
	consider $x$ having $1$ in every left child in that zone and in its root; and take the minimum of ranks of such trees~$x$, over all choices of zones
	(not including the topmost zone, above the first $\mathsf{c}$ on a~branch).
	Thus, every witness of a~tree~$t$ has rank at least $\eta$ if and only if every such zone results in rank at least $\eta$, and the former case does not hold.
	Such a~tree exists for every $\eta<\omega_1$, so the closure ordinal of $\CA$ is $\omega_1$.
	The same holds for layered depths, so the layered ordinal of $\CA$ is $\omega_1$ as well.
\end{exa}

\section{The dichotomy game}

We now move to the definition of the game $\CG^N_\CA$ designed to decide the dichotomy claimed in \cref{thm:main}, in the form stated in \cref{pro:main-layers}.
Once the game is defined, \cref{sec:intuitions} provides intuitions about its structure.
Very broadly, the game can be understood as a~variant of a~\emph{universality game}: given a~non\=/deterministic parity tree automaton, check if it accepts all trees.
The most direct approach to this problem is to construct the complement automaton and play the emptiness game for it.
Another way is directly designing a~game similar to the one in Hausmann and Piterman~\cite[Section~4.2]{Hausmann2022}:
one player constructs a~tree and shows that all runs of the given automaton reject it, while the opponent's role boils down to choosing a~single branch of this tree, which is then analysed.
Noticeably, the combinatorial structure of both games mentioned above is equivalent, if we take into account how the complement automaton is constructed.
The crucial role in these games is played by \emph{selecting functions}, that is, functions assigning directions $d\in\{\dL,\dR\}$ to transitions of the automaton;
they are played in order to show that all runs of the automaton reject a~tree.
The game involved here has a~similar structure, however it is a~bit more convoluted, because it also takes into account ranks of the witness trees $x$.

Note first that, for every regular relation $\Gamma\subseteq\trees_{A\times\{0,1\}}$, there exists another regular relation $\Gamma'\subseteq\trees_{A\times\{0,1\}}$ with the same closure and layered ordinals,
but such that $\pi_A(\Gamma')$ is the set of all trees over $A$.
To see this, it is enough to take
\[\Gamma'=\Gamma\cup\big\{(t,x')\mid x'\in\WF\land\neg\exists x.\,(t,x)\in\Gamma\big\}.\]
Then $\min_{x\in \Gamma_t}\,\rankR(x)=\min_{x\in \Gamma'_t}\,\rankR(x)$ for $t\in\pi_A(\Gamma)$ and $\min_{x\in \Gamma'_t}\,\rankR(x)=0$ for $t\not\in\pi_A(\Gamma)$.

Towards the proof of \cref{pro:main-layers}, we consider some relation $\Gamma\subseteq\trees_{A\times\{0,1\}}$ such that $\Gamma\subseteq\trees_A\times\WF$ and $\pi_A(\Gamma)=\trees_A$.
We assume that $\Gamma$ is given by a~pruned (i.e.,~every state appears in some accepting run) automaton $\CA=(A,Q, q_{\init}, \Delta,\Omega)$.
Reformulating \cref{pro:main-layers} slightly, we need to show that either $\limitF(\Gamma)< N$ for some $N\in\N$, or $\limitR(\Gamma)=\omega_1$, and we can effectively decide which case holds.
We additionally show that if $\limitF(\Gamma)<\omega$ then the value $\limitF(\Gamma)\in \N$ is precisely characterised by the proposed game.

First, a~\emph{side} is a~symbol $s\in\{\sR,\sT\}$ (which stands for ``reach'' and ``trunk''),
and a~\emph{mode} is a~symbol $m\in\{\fg,\fb\}$ (which stands for non\=/branching and binary\=/branching).

A~\emph{state\=/flow} is a~triple $\big((q,s), m, (q',s')\big)$, where $q,q'\in Q$, $s,s'\in\{\sR,\sT\}$, and $m\in\{\fg,\fb\}$.
When $s\neq s'$, we sometimes say this state\=/flow \emph{changes sides from $s$ to $s'$}.
A~\emph{flow} $\mu$ is a~set of state\=/flows.
The set $\big\{(q',s')\mid\big((q,s),m,(q',s')\big)\in\mu\big\}$ is called the \emph{image} of $\mu$.
Note that the number of all possible state\=/flows, hence also of all possible flows, is finite.

We say that a~flow $\bar\mu$ is a~\emph{back\=/marking} if for every pair $(q',s')$ in the image of $\bar\mu$, precisely one state\=/flow leading to $(q',s')$ belongs to $\bar{\mu}$.
This state flow is called \emph{back\=/marked} for $(q',s')$.

Given a~sequence of flows, $\mu_1,\mu_2,\dots$ we can define their \emph{composition} as the graph with vertices $(q,s,n)\in Q\times\{\sR,\sT\}\times\N$
and with a~directed edge from $(q,s,n)$ to $(q',s',n{+}1)$ labelled by $m$ for every state\=/flow $\big((q,s),m,(q',s')\big)\in\mu_{n+1}$, $n\in\N$.
Note that in a~flow there may be two state\=/flows with the same pairs $(q,s)$ and $(q',s')$, with modes $m=\fg$ and $m=\fb$, leading to two parallel edges in this graph.

A~quadruple $(\delta,s,m,d)\in\Delta\times\{\sR,\sT\}\times\{\fg,\fb\}\times\{\dL,\dR\}$ such that $s=\sR$ implies $m=\fg$ (i.e.,~$m=\fb$ is allowed only for $s=\sT$)
is called a~\emph{selector for $(\delta,s)$}.
Such a~selector \emph{agrees with} a~direction $d'\in\{\dL,\dR\}$ if either $m=\fb$ or $d'=d$ (i.e.,~both directions $d'$ are fine if $m=\fb$, but if $m=\fg$ then we require $d'=d$).

Suppose now that we have a~selector $(\delta,s,m,d)$ that agrees with a~direction $d'$, where $\delta=(q,(a,i),q_\dL,q_\dR)$.
The \emph{output side} $s'$ of this selector in direction $d'$ is
\begin{itemize}
\item	$s'\eqdef\sR$ if $s=\sR$ and $i=0$, or $s=\sT$ and $d'\neq d$, and
\item	$s'\eqdef\sT$ otherwise: if $s=\sR$ and $i=1$, or $s=\sT$ and $d'=d$.
\end{itemize}
Then, the state\=/flow \emph{sent in direction} $d'$ by this selector is $\big((q,s),m,(q_{d'},s')\big)$ (note that $q_{d'}$ is the state sent by $\delta$ in direction $d'$).

\begin{figure}
\begin{center}
\begin{tikzpicture}

\eval{\ssx}{2.5}
\eval{\ssy}{-1.5}
\eval{\sss}{4.0}

\node[anchor=mid] at (0*\sss, 0.85) {$d=\dL$};
\node[anchor=mid] at (1*\sss, 0.85) {$d=\dR$};

\draw   (0*\sss-0.5*\ssx,0.2) --
        (0*\sss-0.5*\ssx,0.6) -- 
        (0*\sss+0.5*\ssx,0.6) -- 
        (0*\sss+0.5*\ssx,0.2);

\draw   (1*\sss-0.5*\ssx,0.2) --
        (1*\sss-0.5*\ssx,0.6) -- 
        (1*\sss+0.5*\ssx,0.6) -- 
        (1*\sss+0.5*\ssx,0.2);


\node[anchor=mid east] at (-\ssx, 0.5*\ssy-0.00*\sss) {$m=\fg$};
\node[anchor=mid east] at (-\ssx, 0.5*\ssy-0.75*\sss) {$m=\fb$};

\gDot{BL}{(0.0*\ssx+0*\sss,0*\ssy-0.00*\sss)}
\gDot{BR}{(0.0*\ssx+1*\sss,0*\ssy-0.00*\sss)}

\gDot{GL}{(0.0*\ssx+0*\sss,0*\ssy-0.75*\sss)}
\gDot{GR}{(0.0*\ssx+1*\sss,0*\ssy-0.75*\sss)}


\gTrl{BL}{-1}{$(a,i)$}{}
\gTrl{BR}{+1}{$(a,i)$}{}

\gTtb{GL}{-1}{$(a,0)$}{}
\gTtb{GR}{+1}{$(a,0)$}{}

\end{tikzpicture}
\end{center}
\caption{Depiction of the four types of selectors}
\label{fig:selectors}
\end{figure}

\Cref{fig:selectors} depicts the four types of selectors for $(\delta, s)$ where $\delta$ is a~transition and $s\in\{\sR,\sT\}$ is a~side. The selectors depend on the direction $d\in\{\dL,\dR\}$ and mode $m\in\{\fg,\fb\}$. The selectors with mode $m=\fg$ (top row) agree only with the chosen direction (force this direction), while those with mode $m=\fb$ (bottom row) agree with both directions (allow branching). The latter mode is available only when $s=\sT$. The output side of the selector is chosen as follows. If $s=\sR$ then only the top row is possible and the next side $s'$ is $\sR$ if $i=0$ and is $\sT$ if $i=1$. If $s=\sT$ then both modes are possible, with $s'=\sT$ in all cases except when $m=\fb$ and we move in the direction opposite to $d$ (depicted by continuous, non\=/bold lines).

Assume that we are given a~set $Q'\subseteq Q$ and a~letter $a\in A$.
By $\Delta_a(Q')$ we denote the set of transitions of the form $\big(q, (a,i), q_\dL,q_\dR\big)$ with $q\in Q'$.

Finally, for two sets of states $R,T\subseteq Q$ we denote $\pair{R,T}\eqdef(R{\times}\{\sR\})\cup(T{\times}\{\sT\})$.
Note that every subset of $Q\times\{\sR,\sT\}$ can be uniquely represented as $\pair{R,T}$ for some $R,T\subseteq Q$.

We can now move to the definition of the crucial game used to prove the desired dichotomy.
The game, denoted $\CG^N_\CA$, is parametrised by a~value $N\in\N\cup\{\infty\}$---either a~natural number or a~formal symbol ``infinity''.

The positions of $\CG^N_\CA$ are of the form $\pair{R,T}\subseteq Q\times\{\sR,\sT\}$
(formally, one also needs additional auxiliary positions to represent the situation between particular steps of a~round; we do not refer to these positions explicitly).
The initial position is $\pair{\{q_\init\}, \emptyset}$, where $q_\init$ is the initial state of the automaton $\CA$.
The consecutive steps in a~round $n\in\N$ from a~position $\pair{R_n,T_n}$ are as follows:
\begin{enumerate}
\item \label{st:G-letter}	$\eve$ declares a~letter $a_n\in A$.
\item \label{st:G-selectors}	$\eve$ declares a~set $F_n$ of selectors, containing one selector for each $(\delta,s)\in(\Delta_{a_n}(R_n){\times}\{\sR\})\cup(\Delta_{a_n}(T_n){\times}\{\sT\})$.
\item \label{st:G-direction}	$\adam$ declares a~direction $d_{n+1}\in\{\dL,\dR\}$.
\item \label{st:G-flow}	We define a~flow $\mu_{n+1}$ as the set containing the state\=/flows sent in direction $d_{n+1}$ by all selectors from $F_n$ that agree with this direction.\label{it:step-flow}
\item \label{st:G-backmarking}	$\adam$ declares some back\=/marking $\bar\mu_{n+1}\subseteq\mu_{n+1}$.
\item \label{st:G-image}	We take the image of the back\=/marking $\bar\mu_{n+1}$ as a~new position of the game, $\pair{R_{n+1},T_{n+1}}$.
\end{enumerate}
\noindent 
Given a~play $\Pi$ of the game, the winning condition for $\eve$ is the disjunction \itemfont{A($N$)}${}\lor{}$\itemfont{B} of the following parts.
\begin{description}[leftmargin=4em]
\item[A($N$)] In the graph obtained as a~composition of the back\=/markings $\bar\mu_1,\bar\mu_2,\dots$ there exists a~path which changes sides from $\sR$ to $\sT$ at least $N$ times.%
	\footnote{If $N$ is infinity, then the path should change sides infinitely many times; otherwise, the concrete value of $N$ is hard-coded into the winning condition of the game.}
\item[B] In the graph obtained as a~composition of the flows $\mu_1,\mu_2,\dots$ every infinite path either is rejecting or contains infinitely many state\=/flows of mode $\fb$.
\end{description}
\noindent 
Above, while saying that a~path going through $(q_0,s_0,0),(q_1,s_1,1),\dots$ is rejecting, we mean that the sequence of states $(q_n)_{n\in\N}$ is rejecting, that is, $\limsup_{n\to\infty}\Omega(q_n)$ is odd.
Note that part \itemfont{B} can be restated using an~implication: for every infinite path, if the path is accepting, then it contains infinitely many state\=/flows of mode $\fb$.

\begin{rem}
\label{rem:decide-winner}
	The arena of the game $\CG^N_\CA$ is finite and the winning condition defined above is $\omega$\=/regular.
	Thus, the theorem of Büchi and Landweber~\cite{buchi_synthesis} applies: one can effectively decide the winner of $\CG^N_\CA$.
	Moreover, there exists a~computable bound $M$ such that whoever wins~$\CG^N_\CA$, can win using a~finite\=/memory strategy that uses at most $M$ memory states.
\end{rem}

The following proposition formalises the relation between $\CG^N_\CA$ and \cref{pro:main-layers}.
It holds under the assumption that $\CA$ is a~pruned automaton which recognises a~relation $\Gamma\subseteq\trees_A\times\WF$ such that $\pi_A(\Gamma)=\trees_A$ (i.e.,~the projection of $\Gamma$ is full).
As already explained, this can be assumed without loss of generality.

\begin{prop}
	Under the above assumptions:
	\begin{enumerate}
	\item if player \eve wins $\CG^\infty_\CA$ then $\limitF(\Gamma)=\omega_1$;
	\item if player \adam wins $\CG^\infty_\CA$ then there exists $N\in\N$ such that \adam wins $\CG^N_\CA$;
	\item for any $N\in\N$, player \eve wins $\CG^N_\CA$ if and only if $\limitF(\Gamma)\geq N$.
	\end{enumerate}
\end{prop}
\noindent 
The proof of this proposition is split across the following lemmata.

\begin{restatable}{lem}{lemSoundness}
\label{lem:soundness}
	If player \eve wins $\CG^N_\CA$ then
	\begin{itemize}
		\item $\limitF(\Gamma)=\omega_1$ if $N=\infty$;
		\item $\limitF(\Gamma)\geq N$ if $N\in\N$.
	\end{itemize}
\end{restatable}

\begin{restatable}{lem}{lemCompleteness}
\label{lem:completeness}
	If $\limitF(\Gamma)\geq N$ for some $N\in\N$ then player \eve wins $\CG^N_\CA$.
\end{restatable}

\begin{lem}\label{lem:finitary}
	If player \adam wins $\CG^\infty_\CA$ then there exists $N\in\N$ such that \adam wins $\CG^N_\CA$.
\end{lem}
\noindent 
The first two lemmata are proven in the subsequent sections: soundness (\cref{lem:soundness}) in \cref{sec:soundness}
and completeness (\cref{lem:completeness}) in \cref{sec:completeness}. \cref{lem:finitary} is proven below.

\begin{proofof}{\cref{lem:finitary}.}
	Recall that \cref{rem:decide-winner} gives a~bound $M$ on the number of memory states needed by either of the players to win the game $\CG^\infty_\CA$.
	As $N$ we take the number of positions in the game~$\CG^\infty_\CA$, times $M$, times the number of states of $\CA$, plus one.
	
	Assume that \adam wins $\CG^\infty_\CA$ and fix his finite\=/memory strategy that uses at most $M$ memory states.
	Our goal is to show that the same strategy wins $\CG^N_\CA$.
	Assume for the sake of contradiction that there exists a~play $\Pi$ consistent with this strategy, which makes \eve win $\CG^N_\CA$.
	Since $\Pi$ is losing for \eve in $\CG^\infty_\CA$ it means that
	some back\=/marked path in $\Pi$ contains at least $N$ changes of sides from~$\sR$ to $\sT$.
	Consider the positions just after every such change of sides.
	By the pigeonhole principle, we can find a~repetition among them:
	two moments with the same position of the game, the same memory state, and the same state of $\CA$ on the considered path
	(located on the $\sT$ side, just after a~change of sides).
	We can then repeat forever the fragment of $\Pi$ between these two moments, obtaining another play, also consistent with the fixed strategy of \adam.
	Inside this play we have repetitions of a~fragment of our path, which compose together into an~infinite back\=/marked path with infinitely many changes of sides.
	It follows that the play is winning for \eve in $\CG^\infty_\CA$ by condition \itemfont{A($\infty$)}, contrary to the fact that it is consistent with a~winning strategy of \adam.
\end{proofof}

\begin{proofof}{\cref{pro:main-layers} and \cref{thm:main}.}
	To prove \cref{pro:main-layers} (which implies \cref{thm:main}) we combine the above lemmata.
	First of all we solve $\CG^\infty_\CA$.
	If \eve wins then $\limitF(\Gamma)=\omega_1$.
	Otherwise \adam wins $\CG^N_\CA$ for some $N\in\N$.
	Thus, $\limitF(\Gamma)<N$.
	Note that \eve always wins $\CG^0_\CA$ and look for the greatest $N_0$ such that \eve wins $\CG^{N_0}_\CA$. Then $\limitF(\Gamma)=N_0$.
\end{proofof}

\begin{figure}
\begin{center}
\begin{tikzpicture}

\eval{\ssx}{1.5}
\eval{\ssy}{-1.25}
\eval{\sss}{1.25}


\node[anchor=mid] at (2*\ssx, 0.85) {$R_n$};
\node[anchor=mid] at (5*\ssx+\sss, 0.85) {$T_n$};

\draw (1*\ssx-0.4,0.2) -- (1*\ssx-0.4,0.6) -- (3*\ssx+0.4,0.6) -- (3*\ssx+0.4,0.2);
\draw (4*\ssx-0.4+\sss,0.2) -- (4*\ssx-0.4+\sss,0.6) -- (6*\ssx+0.4+\sss,0.6) -- (6*\ssx+0.4+\sss,0.2);

\gDot{R00}{(1*\ssx,0*\ssy)}
\gDot{R01}{(2*\ssx,0*\ssy)}
\gDot{R02}{(3*\ssx,0*\ssy)}

\gStT{1}{0}{1}{$q_0$}
\gStT{2}{0}{1}{$q_1$}
\gStT{3}{0}{1}{$q_2$}

\gDot{T00}{(4.0*\ssx+\sss,0*\ssy)}
\gDot{T01}{(6.0*\ssx+\sss,0*\ssy)}

\gStR{4}{0}{1}{$q_3$}
\gStR{6}{0}{1}{$q_4$}


\gDot{R20}{(1.0*\ssx,1*\ssy)}
\gDot{R21}{(2.0*\ssx,1*\ssy)}
\gDot{R22}{(3.0*\ssx,1*\ssy)}

\gDot{T20}{(4.0*\ssx+\sss,1*\ssy)}
\gDot{T21}{(5.0*\ssx+\sss,1*\ssy)}
\gDot{T22}{(6.0*\ssx+\sss,1*\ssy)}

\gLab{$\eve$}{$a_n=\mathsf{a}$}{0}

\gCon{R00}{R20}
\gCon{R01}{R21}
\gCon{R02}{R22}
\gCon{T00}{T20}
\gCon{T00}{T21}
\gCon{T01}{T22}


\gTrl{R20}{-1}{$(a,0)$}{0}
\gTrl{R21}{-1}{$(a,1)$}{1}
\gTrl{R22}{+1}{$(a,0)$}{2}

\gTtb{T20}{+1}{$(a,0)$}{3}
\gTtl{T21}{-1}{$(a,1)$}{4}
\gTtb{T22}{-1}{$(a,0)$}{5}

\gLab{$\eve$}{$F_n$}{1}


\gDot{R30}{(1.5*\ssx,3*\ssy)}
\gDot{R31}{(2.5*\ssx,3*\ssy)}

\gStT{1.5}{3}{-1}{$q'_0$}

\gDot{T30}{(4.5*\ssx+\sss,3*\ssy)}
\gDot{T311}{(5.4*\ssx+\sss,3*\ssy)}
\gDot{T312}{(5.6*\ssx+\sss,3*\ssy)}

\gLab{$\adam$}{$d_{n+1}=\dL$}{2}

\gCon{R20-T}{R30}
\gCon{T20-R}{R31}
\gCon{R21-T}{T30}
\gCon{T21-T}{T311}
\gCon{T22-T}{T312}


\gDot{R41}{(2.5*\ssx,4*\ssy)}

\gStT{2.5}{4}{-1}{$q'_1$}

\gDot{T40}{(4.5*\ssx+\sss,4*\ssy)}
\gDot{T41}{(5.5*\ssx+\sss,4*\ssy)}

\gStR{4.5}{4}{-1}{$q'_1$}
\gStR{5.5}{4}{-1}{$q'_2$}

\gLab{$\adam$}{$\mu_{n+1}$}{3}

\gCon{R31}{R41}
\gCon{T30}{T40}
\gCon{T311}{T41}
\gConP{T312}{T41}


\node[anchor=mid] at (2.5*\ssx, 4*\ssy-0.85) {$R_{n+1}$};
\node[anchor=mid] at (5*\ssx+\sss, 4*\ssy-0.85) {$T_{n+1}$};

\draw (2*\ssx-0.4,4*\ssy-0.2) -- (2*\ssx-0.4,4*\ssy-0.6) -- (3*\ssx+0.4,4*\ssy-0.6) -- (3*\ssx+0.4,4*\ssy-0.2);
\draw (4*\ssx-0.4+\sss,4*\ssy-0.2) -- (4*\ssx-0.4+\sss,4*\ssy-0.6) -- (6*\ssx+0.4+\sss,4*\ssy-0.6) -- (6*\ssx+0.4+\sss,4*\ssy-0.2);
\end{tikzpicture}
\end{center}
\caption{A~depiction of a~round of the game $\CG^N_\CA$}
\label{fig:round-of-game}
\end{figure}

\Cref{fig:round-of-game} depicts an~example of a~round of the game $\CG^N_\CA$.
Once the letter $a_n=\mathsf{a}$ is chosen by \eve, there are six possible transitions from the considered states:
one from each of them, except the state $q_3$ which has two possible transitions, $\delta_3$ over $(a,0)$ and $\delta_4$ over $(a,1)$.
Then, \eve declares a~set of selectors
\begin{align*}
	F_n=\Big\{&\big(\delta_0,\sR,\fg,\dL\big),\big(\delta_1,\sR,\fg,\dL\big),\big(\delta_2,\sR,\fg,\dR\big),
	\big(\delta_3,\sT,\fb,\dR\big),\big(\delta_4,\sT,\fg,\dL\big),\big(\delta_5,\sT,\fb,\dL\big)\Big\}.
\end{align*}
Thus, the selectors for the transitions $\delta_3$ and $\delta_5$ are in the mode $\fb$, while the remaining selectors are in the mode~$\fg$.
Once \adam chooses the~direction $d_{n+1}=\dL$, we gather into a~flow $\mu_{n+1}$ the state\=/flows sent in the direction $\dL$ by the selectors corresponding to particular transitions.
The transition $\delta_0$ provides its left state $q_0'$ to the side $\sR$, because this transition is over $(a,0)$.
The transition $\delta_1$ provides its left state $q_1'$ to the side $\sT$, because this transition is over $(a,1)$.
However, $\delta_2$ does not provide its state anywhere, because the direction of the selector is $\dR$ while the direction chosen by \adam is $\dL$ (and the mode is $\fg$, as for all transitions on the side $\sR$).
Coming now to the $\sT$ side: in the case of transition $\delta_3$, the direction $\dR$ of the selector is different from the direction $\dL$ chosen by \adam and therefore the left state of this transition goes to the side $\sR$.
Finally, the transitions $\delta_4$ and $\delta_5$ provide their left states to the side $\sT$; these states are both $q'_2$.
This means that the final flow $\mu_{n+1}$ is
\begin{align*}
	\Big\{\big((q_0,\sR),\fg,(q'_0,\sR)\big),\big((q_1,\sR),\fg,(q'_1,\sT)\big),\big((q_3,\sT),\fb,(q'_1,\sR)\big),\big((q_3,\sT),\fg,(q'_2,\sT)\big),&\\
	&\hspace{-7em}\big((q_4,\sT),\fb,(q'_2,\sT)\big)\Big\}.
\end{align*}
Next, \adam chooses a~back\=/marking $\bar\mu_{n+1}\subseteq\mu_{n+1}$, taking the second, the third, and the fourth state\=/flow from the above set,
which determines the sets $R_{n+1}$ and $T_{n+1}$ constituting the next position of the game.
He had to choose at most one state\=/flow leading to the pair $(q'_2,\sT)$.
The other state\=/flows could be either taken to $\bar\mu_{n+1}$ or not;
in our case \adam decided not to take the state flow leading to $(q'_0,\sR)$, so $q_0'$ is not included in $R_{n+1}$.
Note that some states (in our case $q_1'$) might repeat on both sides.
When resolving the winning condition, part \itemfont{A($N$)} takes into account the back\=/marking $\bar\mu_{n+1}$ chosen by $\adam$, while part \itemfont{B} the full flow $\mu_{n+1}$.

\section{Intuitions}
\label{sec:intuitions}

Let us explain the intuitions behind the games $\CG^N_\CA$, assuming first that $N\in\N$.
Recall the intention: the game is designed so~that \eve wins precisely when there are trees $t$ requiring witnesses $x$ of layered depth at least $N$.

Having this intention in mind, let us discuss details of the game.
The role of \eve should be to show a~tree $t$ (whose all witnesses $x$ have layered depth at least $N$), so we allow her to propose a~label of a~node in step~\ref{st:G-letter}.
However, as usually in games, we do not continue in both children of the current node, but we rather ask \adam to choose one direction ($d_{n+1}$ in step~\ref{st:G-direction}),
where \eve has to continue creating the tree, and where \adam thinks that it is impossible to continue.

Observe now that in a~tree $x$ of layered depth at least $N$, we can always find a~\emph{comb}:
a~node labelled by $1$, followed by an~infinite branch---a~trunk---such that in infinitely many nodes on the side of this trunk the layered depth is at least $N{-}1$.
In~these places we can repeat this process, that is, again find an~analogous comb (located below a~$1$\=/labelled node), and so on, obtaining a~tree of combs nested $N$ times.
Obviously the converse holds as well: such a~tree of combs nested $N$ times can exist inside $x$ only if $x$ has layered depth at least~$N$.

Thus, in order to show that the constructed tree $t$ allows only witnesses $x$ of layered depth at least $N$, an~additional role of \eve should be to show a~nested comb structure.
But this has to be done for every $x$ such that $(t,x)\in\Gamma$, so, in a~sense, for every run of $\CA$ over $(t,x)$ for some $x$.
As usual, we cannot require from \adam to choose a~run on the fly, during the game;
an~interesting (even an~accepting) run of \CA can only be fixed after the whole tree (i.e.,~the whole future of the play) is fixed.
This means that during the game we have to trace all possible runs of $\CA$.
However, there are infinitely many runs, so we cannot do that directly.
To deal with that, we keep track of all ``interesting'' states in the sets $R_n$, $T_n$,
and we consider all possible transitions from them in step~\ref{st:G-selectors}.
This makes the situation of \eve a~bit worse: she has to make decisions based only on the current state, not knowing the past of the run
(the same state may emerge after two different run prefixes).
But it turns out that \eve can handle that; in our proof this corresponds to positionality of strategies in an~auxiliary game considered in \cref{sec:aux-game}.

Now, how exactly does \eve show the nested comb structure?
This is done via selectors proposed in the sets $F_n$.
For states in $R_n$, the role of \eve is to show a~direction in which we can reach a~node with label~$1$.
If \adam follows this direction, and the label of $x$ is $0$, we continue searching for label~$1$, staying on the $\sR$ side
(and if he chooses the opposite direction, we just stop tracing this run).
When label~$1$ is found, we put the resulting state on the $\sT$ side, where \eve shows a~trunk with new combs below it.
To this end, for states in $T_n$ (i.e.,~on the ``trunk'' side) \eve has to show in which direction the trunk continues.
Moreover, \eve has to show places where on the side of the trunk we can find a~nested comb; in these places \eve plays mode~$\fb$ (``branch'').
If \adam chooses a~direction in which the trunk (as declared by~\eve) continues, we trace the resulting state again on the $\sT$ side.
Conversely, if \adam chooses the non\=/trunk direction while the mode is $\fb$, the resulting state ends up on the $\sR$ (``reach'') side.
Then, the role of~\eve is again to find a~node with label $1$ starting the next comb.
The \itemfont{B} part of the winning condition ensures that, for every accepting run, the trunk of each comb has infinitely many branching points (i.e., points with mode $\fb$).
It additionally implies that
on the $\sR$ side we stay only for finitely many steps (as mode $\fb$ does not occur there).
Note that by arbitrarily composing transitions we may obtain also rejecting runs; the \itemfont{B} condition does not require anything for them.

There is one more issue taken into account in the design of the game: \eve should be obliged to produce the combs nested $N$ times, but not more.
The number of \emph{nestings} (i.e.,~of switches between sides $\sR$ and $\sT$) is controlled by \adam.
When he is satisfied with the nesting depth provided by \eve for runs ending in some state,
he can remove this state from the next position in step~\ref{st:G-backmarking} (by providing a~back\=/marking $\bar\mu_{n+1}$ with a~smaller image than that of the flow $\mu_{n+1}$),
and let \eve provide appropriate comb structures only from remaining states.
The \itemfont{A($N$)} part of the winning condition obliges \adam to indeed remove a~state after seeing $N$ nestings.
It is somehow important that \itemfont{A($N$)} talks only about a~back\=/marked history of a~state, not about all possible run prefixes leading to it (i.e., about a~composition of full flows).
Indeed, consider a~situation with two runs leading to some state:
one with already $N{-}1$ nestings of combs provided by \eve, and another where we are still on the trunk of the first comb.
Because \eve should provide $N$ nested combs for all runs, in such a~situation \adam should still be able to request this for the latter run.
To this end, he can select the latter run as the back\=/marked history of the considered state, and continue waiting for $N$ further nested combs.

The situation in $\CG^\infty_\CA$ is essentially the same, except that there is no fixed bound $N$ for the number of nestings of combs that have to be provided by \eve.
Instead, \adam can request more and more nestings of combs on different branches of the tree (but only finitely many on every branch, due to the \itemfont{A($\infty$)} condition);
effectively, this means that \adam can request nested comb structures of arbitrarily large layered depths.
Thus---by design of the game---\adam~wins if there is any countable bound $\eta$ such that every tree $t$ has a~witness $x$ of layered depth at most $\eta$, and \eve wins otherwise.
But because this is a~finite game with an $\omega$\=/regular winning condition, if \adam wins then he wins with a~finite\=/memory winning strategy;
from such a~strategy we can deduce that the bound $\eta$ is actually a~natural number $N$ depending on the size of the memory of $\adam$.
This switch to finite\=/memory strategies (together with determinacy of the game: either \eve wins, or \adam wins with finite memory) underlies the obtained dichotomy:
either there are trees requiring witnesses of arbitrarily large countable layered depth, or there is a~bound $N\in\N$ on layered depths of the witnesses that are needed.

\begin{exa}
	Let us see how the game $\CG^\infty_\CA$ behaves for the automaton $\CA$ from \cref{ex:automaton}.
	Recall that the closure ordinal of $\CA$ is $\omega_1$, so \eve should be able to win.
	
	The strategy of \eve from a~position $\pair{R_n,T_n}$ is as follows.
	If $R_n\cup T_n$ contains a~state~$r_i$, then \eve plays letter $\mathsf{b}$, otherwise letter $\mathsf{c}$.
	Then \eve proposes selectors:
	transitions $\delta$ originating from states $q_j$ are handled by selectors $(\delta,s,m,d)$ with mode $m=\fg$
	and with direction $d$ being such that $\delta$ sends a~state other than $p_0$ in this direction (i.e., $d=\dL$ for $\delta=(q_j,(a,0),q_2,p_0)$, etc.);
	transitions originating from $r_i$ are handled by direction $d=\dR$, and by mode $m=\fg$ on the $\sR$ side and mode $m=\fb$ on the $\sT$ side.
	As we argue below, $p_0$~never becomes an~element of $R_n\cup T_n$, so transitions from $p_0$ need not be handled.
	
	The initial position is $\pair{\{q_1\},\emptyset}$.
	Here $\eve$ plays letter $\mathsf{c}$ and some selectors with mode~$\fg$, and \adam chooses a~direction.
	There are two selectors $(\delta,s,m,d)$ that agree with this direction, and they send there states $q_3$ and $r_1$.
	Thus, the next position, reached via a~flow $\{((q_1,\sR),\fg,(q_3,\sR)),((q_1,\sR),\fg,(r_1,\sR))\}$, is $\pair{\{q_3,r_1\},\emptyset}$.
	This time $\eve$ plays letter $\mathsf{b}$.
	Note that the only transition from $r_1$ reading letter $\mathsf{b}$ on the first coordinate, reads $1$ on the second coordinate; \eve chooses direction $\dR$ in the selector for this transition.
	Thus, if \adam goes left, the new position is simply $\pair{\{q_2\},\emptyset}$; it behaves like the initial position, because from $q_1$ and from $q_2$ we have the same transitions.
	If \adam goes right, the new position, reached via a~flow $\{((q_3,\sR),\fg,(q_1,\sR)),((r_1,\sR),\fg,(r_0,\sT))\}$, is $\pair{\{q_1\},\{r_0\}}$.
	From $\pair{\{q_1\},\{r_0\}}$ once again letter $\mathsf{b}$ is played.
	This time, the selector for the transition from state $r_0$ has direction $\dR$ and mode $\fb$.
	If \adam goes right, the new position, reached via a~flow $\{((q_1,\sR),\fg,(q_1,\sR)),((r_0,\sT),\fb,(r_0,\sT))\}$, is once again $\pair{\{q_1\},\{r_0\}}$.
	If \adam goes left, the new position, reached via a~flow $\{((q_1,\sR),\fg,(q_2,\sR)),((r_0,\sT),\fb,(r_1,\sR))\}$, is $\pair{\{q_2,r_1\},\emptyset}$;
	analogous to the already considered position $\pair{\{q_3,r_1\},\emptyset}$.

	It is also possible that \adam erases some state from a~position (i.e., plays a~back\=/marking with a~smaller image than the obtained flow).
	If a~state $r_i$ is erased, we end up in a~position $\pair{\{q_j\},\emptyset}$, being like the initial position.
	We may also have positions $\pair{\{r_1\},\emptyset}$ and $\pair{\emptyset,\{r_0\}}$,
	and flows $\{((r_1,\sR),\fg,(r_0,\sT))\}$, $\{((r_0,\sT),\fb,(r_0,\sT))\}$, $\{((r_0,\sT),\fb,(r_1,\sR))\}$ between them.
	Finally, we may also reach $\pair{\emptyset,\emptyset}$.
	
	Note that in our example there is always only one state\=/flow leading to every pair $(q,s)$;
	in consequence, every infinite path in the composition of flows is also present in the composition of back\=/markings.
	
	Let us now check the winning condition.
	One possibility is that infinitely many letters $\mathsf{c}$ were played.
	In these moments no state $r_i$ was present in the position, so the only infinite path in the composition of flows is the path going through appropriate $q_j$ states.
	But after seeing every $\mathsf{c}$ this state was $q_3$, so the path is rejecting; part \itemfont{B} of the winning condition is satisfied
	(we may also have no infinite path, if the $q_j$ state was removed by \adam, but then condition \itemfont{B} holds even more).
	The opposite case is that from some moment on, only letter~$\mathsf{b}$ was played.
	We then also have an~infinite path going, from some moment, through the $r_i$ states (this path really exists: if \adam removes the state $r_0$, then letter $\mathsf{c}$ is played).
	Note that whenever \adam goes left, this path changes sides from $\sT$ to $\sR$, and in the next round returns back to the $\sT$ side.
	If this happens infinitely often, \eve wins by condition \itemfont{A($\infty$)}.
	Otherwise, from some moment on \adam constantly goes right.
	After that, the path going through the $r_i$ states has all state\=/flows of mode $\fb$,
	and the other path (if exists) remains in state $q_1$, so it is rejecting;
	condition \itemfont{B} is satisfied.
\end{exa}

\section{Soundness}
\label{sec:soundness}

We begin by proving the soundness property of $\CG^N_\CA$.

\lemSoundness*
\noindent 
We prove the lemma by first providing a~generic way of unravelling any strategy $\sigma_\eve$ of \eve in $\CG^N_\CA$ (with $N$ either in $\N$ or $\infty$).
For every countable successor ordinal $\eta$, our goal is to construct a~tree $t$ (denoted $\unravel(\sigma_\eve, \eta)$) obtained by inductively unravelling the strategy $\sigma_\eve$ of \eve.
Then we show that $\rankR(x)\geq\eta$ whenever $(t,x)\in\Gamma$, which implies that $\rankR(\Gamma)\geq \eta$.
Fix any countable successor ordinal $\eta$.

To perform the construction in a~uniform way, we first fix ``approximations'' of ordinals:
for every ordinal $\xi$ such that $0<\xi\leq\eta$, and for every $n\in\N$ let $\apr(\xi,n)$ be an~ordinal such that
\begin{itemize}
\item	$\apr(\xi,n)$ is a~successor ordinal,
\item	$\apr(\xi,n)\leq\xi$,
\item	$\lim_{n\to\infty}\apr(\xi,n)=\xi$,
\item	$\apr(\xi,n)\leq\apr(\xi',n')$ for all $\xi,\xi',n, n'$ when $\xi\leq\xi'$ and $n\leq n'$.
\end{itemize}

\begin{lem}
	There exists a~function $\apr$ as above.
\end{lem}

\begin{proof}
	Enumerate all successor ordinals less than or equal to $\eta$ in some order: $\xi_0,\xi_1,\dots$ (there are countably many),
	possibly with repetitions (which is needed if $\eta\in\N$).
	Define
	\[\apr(\xi,n)\eqdef\max\big(\set{1}\cup\set{\xi_i\mid i\leq n\land\xi_i\leq\xi}\big).\]
	Note that the maximum is taken from a~finite set.
	All the items follow readily from the definition.
\end{proof}

We explore some plays consistent with the strategy $\sigma_\eve$, arising from playing against some particular strategy of \adam that we also inductively define.
During this process, we keep track of
\begin{itemize}
\item the current node $v\in\{\dL,\dR\}^\ast$,
\item the current position $\pair{R_v, T_v}$ of the game,
\item a~mapping $\xi_v$, which assigns some non\=/zero ordinals $\leq \eta$ to all elements of the set $\pair{R_v,T_v}$.
\end{itemize}
We additionally ensure that the mapping $\xi_v$ has the property that for every $q\in R_v$ the value $\xi_v(q,\sR)$ is a~successor ordinal.

Initially $v=\epsilon$, the position $\pair{R_\epsilon, T_\epsilon}=\pair{\{q_\init\},\emptyset}$ is the initial position of the game, and $\xi_\epsilon=\set{(q_\init,\sR) \mapsto \eta}$.
Since $\eta$ is a~fixed successor ordinal, the invariant about $\xi_v$ holds.

Consider a~round of the game, starting in $v$, with $\pair{R_v, T_v}$ and $\xi_v$ already defined.

First, following her strategy, \eve plays a~letter $a_v$; we put it in the label of node $v$ in~$t$ (i.e.,~$t(v)\eqdef a_v$).
Then \eve plays a~set $F_v$ containing selectors for the various transitions of~$\CA$ and the task of \adam is to choose a~direction.
For both directions $d\in\{\dL,\dR\}$ moving to the child $vd$ we trace a~play in which \adam chooses the respective direction~$d$.
Let $\mu_{vd}$ be the flow obtained in step~\ref{it:step-flow} of such a~round.

Recall that for each state\=/flow $\big((q,s),m,(q',s')\big)\in\mu_{vd}$ we have $(q,s)\in\pair{R_v,T_v}$.
To such a~state\=/flow we assign an~ordinal:
\begin{itemize}
\item in the case of $s=\sR$, $s'=\sT$ we take $\xi_v(q,s)-1$, relying on the fact that the ordinals on the $\sR$ side are always successor ordinals;
\item in the case of $s=\sT$, $s'=\sR$ we take the $|v|$\=/th ordinal $\apr(\xi_v(q,s),|v|)$ in the fixed approximation of the ordinal $\xi_v(q,s)$ (recall that $\xi_v(q,s)>0$ by assumption, so this is possible);
\item in all the remaining cases we take the source ordinal $\xi_v(q,s)$.
\end{itemize}
\noindent 
Note that both cases with $s'=\sR$ ensure that ordinals assigned to state\=/flows leading to the $\sR$ side are successor ordinals.

Now, having assigned ordinals to each state\=/flow, \adam defines the back\=/markings in a~way that maximises the ordinal of the back\=/marking leading to each pair $(q',s')$.
Draws are resolved arbitrarily.
However, if for some pair $(q',s')$ this maximal ordinal would be $0$, then \adam does not back-mark any state\=/flow leading to this pair.

Moreover, for each $(q',s')$ in the new position $\pair{R_{vd},T_{vd}}$ (i.e.,~in the image of $\bar{\mu}_{vd}$)
we define the new ordinal $\xi_{vd}(q',s')$ as the maximal ordinal which is assigned to a~state\=/flow leading to $(q',s')$ (the one back\=/marked by \adam).
In particular, this new $\xi_{vd}(q',s')$ is always positive, is a~successor ordinal if $s'=\sR$, and is less\=/or\=/equal to the corresponding $\xi_v(q,s)$.

This inductive process is continued forever; it produces a~full infinite binary tree $t$, that we denote $\unravel(\sigma_\eve, \eta)$.

\begin{clm}
	If $N\in\N\cup\{\infty\}$, the strategy $\sigma_\eve$ wins $\CG^N_\CA$, and $\eta\leq N$ (for $N=\infty$ this gives the trivial inequality $\eta\leq \infty$)
	then condition \itemfont{A($N$)} is never satisfied by a~play obtained in the above way.
\end{clm}

\begin{proof}
	Consider a~play simulated along some branch $\alpha\in\{\dL,\dR\}^\omega$ of $t$.
	Consider also a~path $\pi$ in the composition of back\=/markings created during this play.
	We have to prove that this path does not change sides from $\sR$ to $\sT$ at least $N$ times.
		
	The path starts on some level\footnote{Due to the structure of back\=/markings, it is always possible to extend any path back to the root by following the unique back\=/marked edges. If the extended path does not satisfy \itemfont{A($N$)}, then neither does the original path. Thus, one could assume without loss of generality that $n_0=0$. However, this is not needed for the argument.} $n_0\in\N$.
	For $n\geq n_0$ let $v_n=\alpha\restr_n$ be the $n$\=/th node of our branch, and let $(q_n,s_n)$ be the element of $\pair{R_{v_n},T_{v_n}}$ located on the path $\pi$.
	Recall the sequence of ordinals $\xi_{v_n}(q_n,s_n)$.
	This sequence is non\=/increasing (because of the choice of back\=/markings by \adam) and strictly decreasing whenever $\pi$ changes sides from $\sR$ to $\sT$ (cf.\@ the first bullet above).
	
	For $N=\infty$ the thesis is now immediate: an~ordinal cannot be decreased infinitely many times.
	
	For $N\in\N$ we recall that all the values of $\xi_v$ are positive (state\=/flows on which the ordinal would decrease to $0$ are never taken into the back\=/marking) and not greater than $\eta\leq N$,
	meaning that we could decrease $\xi_{v_n}(q_n,s_n)$ at most $N{-}1$;
	again \itemfont{A($N$)} does not hold.
\end{proof}

From that moment on assume that $\eta\leq N\in\N\cup\{\infty\}$ and $\sigma_\eve$ wins $\CG^N_\CA$, and consider the tree $t=\unravel(\sigma_\eve,\eta)$.
Thus, the above claim guarantees that all the simulated plays satisfy \itemfont{B}.

It remains to prove that every well\=/founded tree $x$ such that $(t,x)\in\Gamma$ has layered depth at least $\eta$ (i.e.,~$\rankR(x)\geq \eta$).
In the case of $N=\infty$, due to arbitrary choice of $\eta$, this implies that $\limitF(\Gamma)=\omega_1$.
If $N\in\N\setminus\{0\}$ then for $\eta=N$ we obtain that $\rankR(x)\geq N$.
If $N=0$, the thesis $\rankR(\Gamma)\geq N$ holds trivially.

Consider such a~tree $x$ and let $\rho$ be an~accepting run of $\CA$ from the initial state over $(t,x)$, witnessing that $(t,x)\in\Gamma$.

\begin{clm}\label{claim:go-down}
	Suppose that $(\rho(v),s)\in\pair{R_v,T_v}$ for a~node $v$ and for some $s\in\set{\sR,\sT}$.
	In such a~situation, there exist $s'\in\set{\sR,\sT}$, $m\in\set{\fg,\fb}$, and $d\in\set{\dL,\dR}$ such that
	\begin{enumerate}
	\item	the flow $\mu_{vd}$ contains the state\=/flow $\big((\rho(v),s),m,(\rho(vd),s')\big)$,
		\label{it:down-flow}
	\item	if $s' = s$ then $(\rho(vd),s')\in\pair{R_{vd},T_{vd}}$ and $\xi_{vd}(\rho(vd),s')\geq\xi_v(\rho(v),s)$,
		\label{it:down-eq}
	\item	if $s'\neq s$ then $s=\sR$, $x(v)=1$, $s'=\sT$, and either $\xi_v(\rho(v),s)=1$, or $(\rho(vd),s')\in\pair{R_{vd},T_{vd}}$ and $\xi_{vd}(\rho(vd),s')\geq\xi_v(\rho(v),s)-1$, and
		\label{it:down-neq}
	\item	if $m=\fb$ then also $\rho(v\bar{d})\in R_{v\bar{d}}$ and $\xi_{v\bar{d}}(\rho(v\bar{d}),\sR)\geq\apr(\xi_v(\rho(v),s),|v|)$.
		\label{it:down-two}
	\end{enumerate}
\end{clm}

\begin{proof}	
	Consider the transition $\delta$ used by $\rho$ in $v$,
	and the selector $(\delta,s,m,d)\in F_v$ declared by \eve for $(\delta,s)$ when the play was in the node $v$.
	Let $s'$ be the output side of this selector in direction $d$.
	Then $\mu_{vd}$ by definition contains $\big((\rho(v),s),m,(\rho(vd),s')\big)$ and we get Item~\ref{it:down-flow}.

	If $s'=s$ then we have assigned the ordinal $\xi_v(\rho(v),s)>0$ to the state\=/flow $\big((\rho(v),s),\allowbreak m,\allowbreak (\rho(vd),s')\big)$
	and then we have chosen $\xi_{vd}(\rho(vd),s')$ as the maximum of ordinals assigned to state\=/flows leading to $(\rho(vd),s')$,
	which implies Item~\ref{it:down-eq} (in particular some state\=/flow leading to $(\rho(vd),s')$ is back-marked, so indeed $(\rho(vd),s')\in\pair{R_{vd},T_{vd}}$).
	
	Similarly, if $s'\neq s$ then due to the definition of the output side of a~selector we know that $s=\sR$, $x(v)=1$, and $s'=\sT$.
	Moreover, we have assigned the ordinal $\xi_v(\rho(v),s)-1$ to the state\=/flow $\big((\rho(v),s),m,(\rho(vd),s')\big)$.
	This implies Item~\ref{it:down-neq} in the same manner as above.
	
	Finally, if $m=\fb$, then $\mu_{v\bar{d}}$ contains the state\=/flow $\big((\rho(v),s),m,(\rho(v\bar{d}),\sR)\big)$ by definition.
	To this state\=/flow we have assigned the ordinal $\apr(\xi_v(\rho(v),s),|v|)>0$,
	and then we have chosen $\xi_{v\bar{d}}(\rho(v\bar{d}),\sR)$ as the maximum of ordinals assigned to state\=/flows leading to $(\rho(v\bar{d}),\sR)$,
	which gives us Item~\ref{it:down-two}.
\end{proof}

Recall that we want to prove that $\rankR(x)\geq\eta$.
Because $\rho(\epsilon)\in R_\epsilon$ and $\xi_\epsilon(\rho(\epsilon),\sR)=\eta$, it is enough to prove the following claim.

\begin{clm}
\label{cl:estimate}
	For every node $v\in\{\dL,\dR\}^\ast$ and $s\in\{\sR,\sT\}$ with $(\rho(v),s)\in\pair{R_v,T_v}$ it holds that
	\begin{equation}
	\label{eq:estimate}
	\xi_v(\rho(v),s)\leq\rankX{s}(x\restr_v).
	\end{equation}
\end{clm}

The rest of this section is devoted to the proof of \cref{cl:estimate}.

\noindent 
Assume for the sake of contradiction that the above conclusion is violated for some pairs $(v,s)$.
Assign to each such pair $(v,s)$ its \emph{weight} $(\gamma, s)$ defined as $(\rankX{s}(x\restr_v),s)$.
We order weights according to the lexicographic order: we first compare the ordinals $\gamma$ and then sides~$s$, with $\sR<\sT$.
Let $(\gamma, s)$ be the minimal weight among all pairs violating the conclusion of \cref{cl:estimate}.

We consider separately the two possible cases for $s$.

\paragraph*{The $\sR$ side}
	First suppose that the minimal weight $(\gamma, s)$ has $s=\sR$.
	Take a~pair $(v,s)$ of this weight.
	Then we have $\xi_v(\rho(v),s)>\rankR(x\restr_v)=\gamma$.
	Due to the assumptions on $\xi_v$ when $s=\sR$ we know that $\xi_v(\rho(v),s)$ is a~successor ordinal.
	
	Let us repeatedly use \cref{claim:go-down} for $w_0=v$ to go down from the current node until we get $s'=\sT$.
	This process results in a~finite or infinite sequence of successive nodes $v=w_0\prec w_1\prec \ldots$.
	We preserve the invariant that for every $w$ in this sequence we have $\xi_{w}(\rho(w),s)\geq\xi_v(\rho(v),\sT)\geq\gamma+1$.
	
	First assume that at some point \cref{claim:go-down} gives us a~pair $(\rho(wd),\sT)$ on the $\sT$ side, in which case we have $x(w)=1$ (cf.\@ Item~\ref{it:down-neq} of \cref{claim:go-down}), so
	\[\gamma=\rankR(x\restr_v)\geq\rankT(x\restr_{wd})+1\]
	due to \cref{lem:fin-oank-to-fank}.
	In particular $\gamma\geq 1$, so $\xi_{w}(\rho(w),s)\geq 2$; by the second part of Item~\ref{it:down-neq} we thus have $\xi_{wd}(\rho(wd),\sT)\geq\xi_{w}(\rho(w),s)-1$.
	Altogether we obtain
	\[\xi_{wd}(\rho(wd),\sT)\geq\xi_{w}(\rho(w),s)-1\geq\gamma>\rankT(x\restr_{wd});\]
	which implies that the conclusion is violated for the pair $(wd, \sT)$.
	This contradicts minimality of $(\gamma,\sR)$, because the weight of $(wd, \sT)$ is $(\rankT(x\restr_{wd}),\sT)$, which is smaller in the chosen order than $(\gamma,\sR)$ (we know that $\rankT(x\restr_{wd})< \gamma$).
	
	Now assume that, while going down inductively applying \cref{claim:go-down}, we stay forever on the $\sR$ side.
	We thus have an~infinite branch $\alpha$ in the tree, together with a~path $\pi$ in the graph obtained by composing the state\=/flows seen in the play simulated along $\alpha$.
	In consecutive nodes $w$ on $\alpha$, the path $\pi$ goes through the pairs $(\rho(w),\sR)$.
	This path does not contain any state\=/flows of mode $\fb$ (mode $\fb$ is possible only on the $\sT$ side),
	and is not rejecting (because the run $\rho$ is accepting), so it violates condition \itemfont{B}; a~contradiction.

\paragraph*{The $\sT$ side}
	Next assume that the minimal weight $(\gamma, s)$ has $s=\sT$.
	Take a~pair $(v,s)$ of this weight.
	Then we have $\xi_v(\rho(v),\sT)>\rankT(x\restr_v)=\gamma$.
	
	Again, apply inductively \cref{claim:go-down} for $w_0=v$ tracking down the run $\rho$ along some path~$\alpha$.
	Since \cref{claim:go-down} ensures that whenever $s=\sT$ then also $s'=\sT$, we know that the effect is an~infinite branch $\alpha\in\{\dL,\dR\}^\omega$ such that $v\prec \alpha$.

	We preserve the invariant that for all nodes $w\succeq v$ on $\alpha$ we have $\rho(w)\in T_w$ and $\xi_w(\rho(w),\sT)\geq\xi_v(\rho(v),\sT)\geq\gamma+1$.
	Moreover, in the graph obtained by composing the state\=/flows $\mu_w$ seen in the play simulated along $\alpha$
	there is a~path $\pi$ going through all the pairs $(\rho(w),\sT)$ for consecutive nodes $w\succeq v$ on $\alpha$.

	Now, as the run $\rho$ is accepting, the path $\pi$ (containing states of $\rho$ along the branch $\alpha$) cannot be rejecting.
	Therefore, condition \itemfont{B} guarantees that infinitely many state\=/flows of mode $\fb$ occur on $\pi$.
	Let $v\preceq w_0\prec w_1 \prec \ldots$ be an~infinite sequence of nodes on $\alpha$ such that while going down from $w_i$ to its child~$w_id_i$ (along $\alpha$) there was indeed a~state\=/flow of mode~$\fb$ on $\pi$.
	Then, by Item~\ref{it:down-two} of \cref{claim:go-down} we know that $\rho(w_i\bar{d_i})\in R_{w_i\bar{d_i}}$ and
	\begin{align*}
		\xi_{w_i\bar{d_i}}(\rho(w_i\bar{d_i}),\sR)\geq\apr\big(\xi_{w_i}(\rho(w_i),\sT),|w_i|\big)\geq\apr(\gamma+1,|w_i|).
	\end{align*}
	Because $\lim_{i\to\infty}\apr(\gamma+1,|w_i|)=\gamma+1$, starting from some $i_0\in\N$ we have $\apr(\gamma+1,|w_i|)=\gamma+1$ for all $i\geq i_0$.
	
	Note that each node $w_i\bar{d_i}$ either satisfies the conclusions of the claim, in which case $\rankR(x\restr_{w_i\bar{d_i}})\geq \xi_{w_i\bar{d_i}}(\rho(w_i\bar{d_i}),\sR)\geq\gamma+1$,
	or it does not satisfy the conclusions, in which case due to minimality of $\gamma$ and the assumption that $\sR<\sT$ we also know that $\rankR(x\restr_{w_i\bar{d_i}})\geq \gamma+1$.

	\cref{lem:inf-fank-to-oank} applied to the nodes $(w_i\bar{d_i})_{i\geq i_0}$ implies that
	$\rankT(x\restr_v)\geq \gamma+1$, contradicting the initial assumption that $\rankT(x\restr_v)=\gamma$.

This concludes the proof of \cref{cl:estimate} and therefore also of \cref{lem:soundness}.

\section{Auxiliary game}
\label{sec:aux-game}

Before moving towards a~proof of completeness, we need to be able to construct reasonable strategies of \eve in $\CG^N_\CA$.
This is achieved by considering an~auxiliary game, based directly on $\CG^N_\CA$, when considering a~single transition of $\CA$ at each round.
The players of the game are called Automaton (responsible for choosing transitions and choices of \adam in $\CG^N_\CA$) and Pathfinder (responsible for choices of \eve in $\CG^N_\CA$).
The game is denoted $\CH_{\CA,t,N}$ and depends on the fixed automaton $\CA$, a~tree $t\in\trees_A$, and a~natural number $N\in\N$.
Positions of $\CH_{\CA,t,N}$ are triples $(v,q,s)\in\set{\dL,\dR}^\ast\times Q\times\{\sR,\sT\}$,
plus some additional auxiliary positions, to which we do not refer explicitly.

For a~position $(v,q,s)$ we define
\begin{align*}
\val_t(v,q,s)&\eqdef\inf \left(\{N\}\cup\big\{\rankX{s}(x)\mid \text{$x\in\WF$, $(t\restr_v, x)$ can be accepted from $q$}\big\}\right).
\end{align*}
Note that this value is always well\=/defined and bounded by $N$.

For every position $(v,q,s)$ such that $\val_t(v,q,s)<N$, define $\wit(v,q,s)$ as a~pair $(x,\rho)$, where $x\in \WF$ satisfies $\rankX{s}(x) = \val_t(v,q,s)$ and $\rho$ is an~accepting run on $(t\restr_v, x)$ from $q$ (i.e.,~$\rho(\epsilon)=q$).
Such a~pair exists by the definition of $\val_t(v,q,s)$ and the fact that $\val_t(v,q,s)<N$.

Formally, we do not distinguish an~initial position of the game, because we consider uniform strategies winning from everywhere.
Intuitively, one should think that the game begins in $(\epsilon, q_\init, \sR)$.

In positions $(v,q,s)$ such that $\val_t(v,q,s)=0$ the game reaches an~\emph{immediate victory} in~which Pathfinder wins.
A~round from a~position $(v,q,s)$ such that $\val_t(v,q,s)>0$ consists of the following steps:
\begin{enumerate}
\item Automaton declares a~transition $\delta=\big(q,(t(v),i),q_\dL,q_\dR\big)$ from the current state $q$ over $(t(v),i)$ for some $i\in\{0,1\}$.
\item Pathfinder declares a~selector $(\delta,s,m,d)$ for $(\delta,s)$.
\item Automaton declares a~direction $d'$ which agrees with the selector.
\end{enumerate}
\noindent 
Let $v'=vd'$, $q'=q_{d'}$, and let $s'$ be the output side of the selector in the direction~$d'$.
Now the following four conditions of \emph{immediate victory} may end the game, making Automaton win (the square brackets give names of these conditions):
\begin{alignat*}{8}
\imwin{\sR}{\sR}\quad&s=\sR &&\ \land\ s'=\sR &&\ \land\ && \val_t(v',q',s')< \val_t(v,q,s),\\
\imwin{\sR}{\sT}\quad&s=\sR &&\ \land\ s'=\sT &&\ \land\ && \val_t(v',q',s')< \val_t(v,q,s)-1,\\
\imwin{\sT}{\sT}\quad&s=\sT &&\ \land\ s'=\sT &&\ \land\ && \val_t(v',q',s')< \val_t(v,q,s),\\
\imwin{\sT}{\sR}\quad&s=\sT &&\ \land\ s'=\sR &&\ \land\ && \val_t(v',q',s')< \val_t(v,q,s).
\end{alignat*}
Note that these conditions are pairwise exclusive, so at most one immediate victory out of these four can happen in a~fixed round of the game.

If no immediate victory happened, the game proceeds to the new position which is $(v',q',s')$.
Note that the conditions of immediate victory depend only on $(v,q,s)$ and $(v',q',s')$, so can be directly hardwired in the structure of the game.
We do it in such a~way that if an~immediate victory condition is satisfied, then the game goes to a~sink position in which the respective player wins.
In particular, the next position $(v',q',s')$ is reached only if no immediate victory happened.

An~infinite play (i.e.,~without any immediate victory) of $\CH_{\CA,t,N}$ is won by Pathfinder if the sequence of visited states is rejecting or a~selector with mode $\fb$ is played infinitely often.

The main result about $\CH_{\CA,t,N}$ is the following lemma.
Note that it does not assume anything about the value $\val_t(\epsilon,q_\init,\sR)$ for the root of $t$.

\begin{lem}\label{lem:pathfinder-wins}
	Pathfinder has a~positional strategy in $\CH_{\CA,t,N}$ that is winning from every position of the game (we call such a~strategy \emph{uniform}).
\end{lem}
\noindent 
The rest of this section is devoted to a~proof of this lemma.

Note first that the winning condition of Pathfinder in $\CH_{\CA,t,N}$ can be written as a~parity condition.
Indeed, it is enough to shift the priorities of states from $\CA$ by one (to swap rejecting with accepting), and assign a~high even priority to moves choosing a~selector of mode $\fb$;
immediate victory can be realised by a~sink node with a~self\=/loop.
Recall also that parity games are positionally determined~\cite{jutla_determinacy,mostowski_parity_games}.
Thus, in order to prove \cref{lem:pathfinder-wins}, it is enough to show that Automaton does not have a~winning strategy from any position of~$\CH_{\CA,t,N}$.

Assume thus for the sake of contradiction that Automaton has a~winning strategy $\sigma_0$ from some position of $\CH_{\CA,t,N}$
(it may be convenient to think about a~positional strategy, but the proof written below does not depend on positionality of $\sigma_0$).

\begin{clm}
\label{cl:no-change-sides}
	There exists a~position $(v_0,q_0,s_0)$ of $\CH_{\CA,t,N}$ reachable by the strategy $\sigma_0$ such that the following property holds ($\clubsuit$):
	\begin{quote}
		In plays consistent with $\sigma_0$ from $(v_0,q_0,s_0)$, whenever we move from a~position $(v,q,s)$ to a~position $(v',q',s')$ using a~selector with a~mode $m$ (without an~immediate victory), then $s=s'$ and $m=\fg$.
	\end{quote}
\end{clm}

\begin{proof}
	If the thesis of the claim does not hold, then from each position reached by a~partial play consistent with $\sigma_0$, Pathfinder can force to reach either a~change of sides,
	or a~move with mode $\fb$.
	In such a~case, Pathfinder can inductively construct a~play that is consistent with $\sigma_0$ and contains infinitely many such events.
	Every second change of sides is from~$\sT$ to~$\sR$, and it can occur only if a~selector with mode $\fb$ was played.
	It follows that our play is won by Pathfinder, because it needs to contain infinitely many selectors with mode $\fb$.
	This contradicts the fact that $\sigma_0$ is winning for Automaton.
\end{proof}

Without loss of generality assume that the initial position $(v_0,q_0,s_0)$ of $\sigma_0$ satisfies Property ($\clubsuit$).

Similarly as in \cref{sec:soundness}, we assign \emph{weights} to positions $(v,q,s)$ reachable by $\sigma_0$.
The \emph{weight} of a~position $(v,q,s)$ is the number $k=\val_t(v,q,s)\in\{0,\ldots,N\}$.

Take a~maximal weight $k_0$ among all positions reachable by $\sigma_0$.
Again, without loss of generality assume that the initial position $(v_0,q_0,s_0)$ of $\sigma_0$ has this maximal weight $k_0$.

First note that $k_0>0$, as otherwise we would have $\val_t(v_0,q_0,s_0)=0$, which means that Pathfinder wins immediately in $(v_0,q_0,s_0)$, contradicting the fact that $\sigma_0$ is winning for Automaton.

\begin{clm}
\label{cl:no-change-weight}
In all positions $(v,q,s)$ reachable by $\sigma_0$ from $(v_0,q_0,s_0)$ without any immediate victory, the weight of $(v,q,s)$ equals $k_0$.
\end{clm}

\begin{proof}
We ensure this equality inductively over the length of the considered play.
Initially the position is $(v_0,q_0,s_0)$ and the weight is $k_0$.

Consider a~round starting in $(v,q,s)$ (of weight $k_0$) and reaching $(v',q',s')$ without any immediate victory.
First of all $s'=s=s_0$ due to Property ($\clubsuit$).

Let $k$ be the weight of $(v',q',s')$.
We need to show that $k=k_0$.
Clearly we have $k\leq k_0$ due to maximality of $k_0$.
Assume for the sake of contradiction that $k<k_0$.
It means that the immediate victory $\imwin{s_0}{s_0}$ happens in this round, contradicting the assumption that no immediate victory happened.
\end{proof}

We now proceed to the final part of the proof of contradiction in \cref{lem:pathfinder-wins}.
Consider the two possible cases of the side $s_0$.

\paragraph*{When $s_0=\sR$.}

We unravel the strategy $\sigma_0$ by considering all the possible counter\=/strategies of Pathfinder:
he has to play $m=\fg$ (due to $s=\sR$) and we consider both possible choices of directions $d\in\{\dL,\dR\}$.
Since $m=\fg$, Automaton can only choose $d'=d$, and then $v'=vd$.
In each step, while being in a~node $v$, the strategy $\sigma_0$ provides us with a~label $i_v\in\{0,1\}$ and a~transition $(q,(t(v),i_v),q_\dL,q_\dR)$ of $\CA$ reading $(t(v),i_v)$.
We collect the labels $i_v$ obtained this way into a~tree $x$ and transitions into a~run $\rho$, that is, we set $x(v)=i_v$ and $\rho(v)=q$.

Note that we can never reach a~position where $\val_t(v,q,s)=0$, as Pathfinder would win there.
Thus, the above process of unravelling $\sigma_0$ proceeds either forever or until one of the remaining four immediate victory conditions is satisfied.

Assume that an~immediate victory $\imwin{\sR}{s'}$ happens in a~position $(v,q,s)$, when $v'=vd$, $q'=q_d$, and $s'$ is the output side of the chosen selector in the direction $d$.

Consider two cases: either $s'=\sR$ or $s'=\sT$.
\begin{itemize}
\item	If $s'=\sR$ then we know that $\val_t(v',q',s')<k_0$. We use the pair $(x_{v'},\rho_{v'})$ given by $\wit(v',q',s')$ to plug its components into the tree $x$ and run $\rho$
	so that $x\restr_{v'}=x_{v'}$ and $\rho\restr_{v'}=\rho_{v'}$.
	This ensures that $\rankR(x\restr_{v'})<k_0$.
\item	If $s'=\sT$ then we know that $\val_t(v',q',s')<k_0{-}1$.
	Again we use $\wit(v',q',s')$ to plug its components into $x$ and $\rho$ under the node $v'$.
	This ensures that $\rankT(x\restr_{v'})<k_0{-}1$, hence also $\rankR(x\restr_{v'})\leq\rankT(x\restr_{v'})+1<k_0$.
\end{itemize}
\noindent 
This whole process gives us a~tree $x$ together with a~run $\rho$ of $\CA$ on $(t\restr_{v_0},x)$.
We first argue that the run is accepting.
Clearly, if a~path goes through a~final node $v'$ then the run is accepting by the choice of $\rho\restr_{v'}$.
Otherwise, the path goes along an~actual play of $\CH_{\CA,t,N}$ which is consistent with $\sigma_0$ and therefore the sequence of states must be accepting,
as otherwise the play would be won by Pathfinder.
This implies that $\val_t(v_0,q_0,s_0)\leq\rankR(x)$.

We claim that $\rankR(x)< k_0$, which contradicts the assumption that $\val_t(v_0,q_0,s_0)= k_0$ following from the choice of $k_0$.

Consider $\derF^{k_0-1}(x)$ (we know that $k_0>0$); we show that it is the empty tree.
Let $u$ be any node which has label $1$ in $x$.
Firstly, assume that $u\succeq v'$ for some final node $v'$ of type $s'$.
Then $\rankR(x\restr_{v'})<k_0$, meaning that $\derF^{k_0-1}(x)\restr_{v'}$ is the empty tree; in particular $\derF^{k_0-1}(x)(u)=0$.

Now assume that $u$ is not a~descendant of any final node $v'$.
Thus, a~position of the form $(u,q,s)$ is reachable from $(v_0,q_0,s_0)$ by $\sigma_0$.
Thus, $s=\sR$ and since $i_u=1$ we know that $s'=\sT$.
This means that both children $ud$ of $u$ are final nodes of type $\sT$, and therefore they satisfy $\rankT(x\restr_{ud})< k_0-1$.
By the definition of $\rankT$ we know that $\rankR(x\restr_u)< k_0$ and again $\derF^{k_0-1}(x)(u)=0$.
Thus we know that $\derF^{k_0-1}(x)$ is indeed the empty tree and $\rankR(x)< k_0$, as needed for contradiction.

\paragraph*{When $s_0=\sT$.}

We again unravel the strategy $\sigma_0$ by considering certain specific counter\=/strategies of Pathfinder.
Consider a~round starting in a~position $(v,q,s)$.
Property ($\clubsuit$) implies that $s=\sT$.
\cref{cl:no-change-weight} implies that $\val_t(v,q,s)=k_0$.

First, Automaton chooses a~transition $\delta=(q,(t(v),i_v),q_\dL,q_\dR)$ including a~label $i_v\in\{0,1\}$.
We set $x(v)=i_v$ and $\rho(v)=q$.
Note that Property ($\clubsuit$) implies that if Pathfinder chooses a~selector with mode $\fb$
then, after Automaton chooses a~direction $d'\in\{\dL,\dR\}$, one of the four immediate victory conditions necessarily applies.

First note that we can always move to any direction $d\in\{\dL,\dR\}$ in the following way, which we call a~$\fg$\=/step in direction $d$:
We examine a~play in which Pathfinder plays the selector $(\delta,s,\fg,d)$.
The automaton has to answer $d'=d$.
If this round leads to an~immediate victory then it must be $\imwin{\sT}{\sT}$, and then $\val_t(vd,q_d,\sT)<k_0$.
In this case we consider $(x_{vd},\rho_{vd})=\wit(vd,q_d,\sT)$ and plug into $x\restr_{vd}$ the tree $x_{vd}$ and into $\rho\restr_{vd}$ the tree $\rho_{vd}$;
the tree satisfies $\rankT(x\restr_{vd})=\val_t(vd,q_{d},\sT)<k_0$.
Otherwise we can continue the unravelling process from position $(vd,q_d,\sT)$, as the whole inductive construction tells us.

However, the $\fg$\=/step is not our preferred way of proceeding.
We rather consider the following three possibilities.

\begin{enumerate}
\item	Suppose first that if Pathfinder plays the selector $(\delta,s,\fb,d)$ with $d=\dL$, then Automaton answers with $d'=\dR\neq d$.
	This round must end with an~immediate victory $\imwin{\sT}{s'}$ because the mode $\fb$ was played.
	Moreover, $s'=\sR$ due to the rules of the game.
	This means that $\val_t(vd',q_{d'},\sR)<k_0$.
	We use the pair $\wit(vd',q_{d'},\sR)$ to plug into $x$ and $\rho$ under $vd'$.
	We obtain $\rankR(x\restr_{vd'})=\val_t(vd',q_{d'},\sR)<k_0$.

	In order to obtain the subtree of $x$ under $vd$ we perform a~$\fg$\=/step in direction~$d$.

\item	Second, suppose that the above does not hold, but if Pathfinder plays the selector $(\delta,s,\fb,d)$ with $d=\dR$, then Automaton answers with $d'=\dL\neq d$.
	Then we proceed symmetrically: in the left subtrees of $x$ and $\rho$ we plug the pair $\wit(vd',q_{d'},\sR)$, so that $\rankR(x\restr_{vd'})=\val_t(vd',q_{d'},\sR)<k_0$,
	and in order to obtain the right subtree of $x$ we perform a~$\fg$\=/step in direction~$d$.

\item	Finally, it is possible that none of the above happened: for both directions $d\in\{\dL,\dR\}$, to $(\delta,s,\fb,d)$ Automaton answers with $d'=d$.
	Both these rounds end in the immediate victory $\imwin{\sT}{\sT}$.
	Then for both directions $d\in\{\dL,\dR\}$ in the respective subtrees of $x$ and $\rho$ we plug the pair $\wit(vd,q_d,\sT)$.
	Since the immediate victory $\imwin{\sT}{\sT}$ happens, we know that $\rankT(x\restr_{vd})=\val_t(t\restr_{vd},q_d,\sT)< k_0$.
\end{enumerate}
\noindent 
Note that the tree $x$ obtained this way consists of a~branch on which we unravel the strategy $\sigma_0$;
the nodes to the sides of this branch are \emph{final}, that is, we plug appropriate witnessing trees from $\wit$ there.
The branch can be infinite, or it can end if an~immediate victory happens in a~$\fg$\=/step, or it can end when the last of the above cases is reached (and then both children are final).
As in the previous case, we can never reach a~position where $\val_t(v,q,s)=0$, as Pathfinder would win there.

Together with $x$, we obtain a~run $\rho$ of $\CA$ over $(t\restr_{v_0},x)$; let us see that this run is accepting on every path.
Clearly, if a~path goes through a~final node then the run is accepting by construction.
Otherwise, the path goes along an~actual play of $\CH_{\CA,t,N}$ which is consistent with $\sigma_0$;
therefore, the sequence of states must be accepting, as otherwise the play would be won by Pathfinder.
This implies that $\val_t(v_0,q_0,\sT)\leq\rankT(x)$.

We claim that $\rankT(x)< k_0$, which contradicts the assumption that $\val_t(v_0,q_0,\sT)=k_0$.
In order to prove this, we analyse the derivative $\derF^{k_0-1}(x)$.

Firstly, for every final node $v'$ of type $\sR$ (i.e., obtained during some of the first two cases of the above construction, outside of $\alpha$) we have $\rankR(x\restr_{v'})<k_0$.
This implies that $\derF^{k_0-1}(x)\restr_{v'}$ is empty.

Secondly, let us observe that label $1$ occurs on $\alpha$ only finitely many times.
To this end, recall that we have an~accepting run $\rho$ of $\CA$ from state $q_0$ over $(t\restr_{v_0},x)$.
But our automaton is pruned, so the state $q_0$ is used in a~node $w$ of some accepting run from the initial state, over some other tree.
In this tree and in the run over it, let us replace the subtrees starting at $w$ by $(t\restr_{v_0},x)$ and $\rho$, respectively.
The resulting run is accepting as well, so the obtained tree belongs to $\Gamma\subseteq\trees_A\times\WF$.
It follows that label $1$ occurs (on the second coordinate) only finitely many times on every branch of the obtained tree;
in particular, it occurs only finitely many times on the branch $\alpha$ in $x$.

Finally, consider final nodes $v'$ of type $\sT$.
If $\alpha$ ended by an~immediate victory during a~$\fg$\=/step, we have precisely one such node;
if $\alpha$ ended when the third case was reached, we have precisely two such nodes;
and if $\alpha$ is infinite, there are no such nodes.
In both cases where those $v'$ exist, we have $\rankT(x\restr_{v'})<k_0$, meaning that $\derF^{k_0-1}(x)\restr_{v'}$ is finite.

Concluding, we obtain that $\derF^{k_0-1}(x)$ contains finitely many labels $1$ and therefore $\rankT(x)<k_0$, as required.

\section{Completeness}
\label{sec:completeness}

We now move to the proof of completeness of $\CG^N_\CA$.

\lemCompleteness*

\noindent 
The condition $\limitF(\Gamma)\geq N$ says that there exists a~tree $t$ for which every witnessing tree $x$ satisfies $\rankR(x)\geq N$.
Fix such a~tree $t$.
The above assumption about $t$ implies that $\val_t(\epsilon,q_\init,\sR)\geq N$.

Based on the tree $t$, we construct a~strategy of \eve in $\CG^N_\CA$.
Together with the current position of the game, $\pair{R_n,T_n}$, we store a~node $v_n$ of the tree, starting with $v_0=\epsilon$.
We need to provide choices of \eve:
she declares a~letter based on the label of $v_n$ in the fixed tree~$t$,
and a~set of selectors based on a~fixed uniform positional strategy $\pi_0$ of Pathfinder in~$\CH_{\CA,t,N}$, given by \cref{lem:pathfinder-wins}.
The current node $v_n$ is then updated according to the direction declared by \adam in $\CG^N_\CA$.

Our goal is to show that the above strategy is winning.
We consider a~play of this strategy, with $(v_n)_{n\in\N}$ and $(\pair{R_n,T_n})_{n\in\N}$ as above.

For a~pair $(q,s)\in\pair{R_n,T_n}$ denote by $\hist_n(q,s)$ the number of switches from the side~$\sR$ to the side~$\sT$ on the back\=/marked history of $(q,s)$.
Note that if $\hist_n(q,s)\geq N$ then the whole play is won by \eve due to \itemfont{A($N$)}, so from that point on assume that $\hist_n(q,s)< N$ for all $(q,s)\in\pair{R_n,T_n}$.

We keep a~\emph{val\=/preserving invariant} saying that for every pair $(q,s)\in\pair{R_n,T_n}$ with $k=N-\hist_n(q,s)$ we have $\val_t(v_n,q,s)\geq k$, where $\val_t$ is defined as at the beginning of \cref{sec:aux-game}.
Note that the val\=/preserving invariant is initially met, because due to our assumptions $\val_t(\epsilon,q_\init,\sR)\geq N$ and $\hist_0(q_\init,\sR)=0$.
Note also that always $k>0$ because we know that $\hist_n(q,s)<N$.

Consider an~$n$\=/th round of the game $\CG^N_\CA$.
First, playing as \eve we declare $t(v_n)$ as the next letter $a_n$.
The selectors played by \eve are chosen according to the fixed strategy $\pi_0$ of Pathfinder.
Namely, in order to obtain a~selector for a~pair $(\delta,s)$, where $\delta=(q,(t(v_n),i),q_\dL,q_\dR)$,
we let Automaton play the transition from position $(v_n,q,s)$ in $\CH_{\CA,t,N}$, and we check the answer of Pathfinder.
As we observed, the val\=/preserving invariant ensures that $k$ is always positive, so we never reach an~immediate victory in which $\val_t(v_n,q,s)=0$.
We cannot reach any of the four immediate victory conditions of Automaton because they make Automaton win.
This means that no immediate victory can happen at all.

After we provided the selectors of \eve, we receive from \adam the next direction $d_{n+1}$ and a~back\=/marking~$\bar{\mu}_{n+1}$ (which we ignore).
We set $v_{n+1}=v_nd_{n+1}$, and we are ready to simulate the next round of the game.

Note that for every state\=/flow $\big((q,s),m,(q',s')\big)$ in the flow $\mu_{n+1}$, the strategy $\pi_0$ in $\CH_{\CA,t,N}$ allows a~move from $(v_n,q,s)$ to $(v_{n+1},q',s')$ using a~selector with mode $m$.
Indeed, this state flow belongs to $\mu_{n+1}$ because \eve played a~selector $(\delta,s,m,d)$ which agrees with direction $d_{n+1}$,
where $q$ is the source state of $\delta$, and $q'$ is the state sent by $\delta$ in direction~$d_{n+1}$.
This means that if Automaton plays $\delta$ from position $(v_n,q,s)$ of $\CH_{\CA,t,N}$, then Pathfinder (according to his fixed positional winning strategy $\pi_0$) answers $(\delta,s,m,d)$,
and then Automaton playing $d_{n+1}$ can go to position $(v_{n+1},q',s')$.
Recall that no immediate victory happens there.

\begin{clm}\label{claim-invariant}
The play defined according to the above rules obeys the val\=/preserving invariant.
\end{clm}

\begin{proof}
	Take a~pair $(q',s')\in\pair{R_{n+1},T_{n+1}}$ and denote $k'=N-\hist_{n+1}(q',s')$.
	Consider the back\=/marked state\=/flow $\big((q,s),m,(q',s')\big)$ leading to $(q',s')$, and denote $k=N-\hist_{n}(q,s)$.
	Note that due to the definition of $\hist_n(q,s)$ (it counts the number of changes of sides from $\sR$ to $\sT$) we have $k'=k$ except the case when $s=\sR$, $s'=\sT$, and $k'=k{-}1$.
	
	As explained above, the state\=/flow $\big((q,s),m,(q',s')\big)$ corresponds to a~move from $(v_n,q,s)$ to $(v_{n+1},q',s')$ in $\CH_{\CA,t,N}$ (with no immediate victory).
	But if the invariant holds for $(q,s)$ and $k$, and it does not hold for $(q',s')$ and $k'$ (i.e.,~$\val_t(v_{n+1},q',s')< k'$),
	then one of the four immediate victory conditions would be satisfied in that move, leading to a~contradiction.
\end{proof}

Consider now the whole infinite play of $\CG^N_\CA$ described above.
We prove that it satisfies part \itemfont{B} of the winning condition.
Consider any infinite path in the graph obtained as the composition of the played flows.
This path corresponds to an~infinite play of $\CH_{\CA,t,N}$ consistent with $\pi_0$, and therefore it is winning for Pathfinder.
What we need from this path for condition \itemfont{B} is exactly the winning condition from $\CH_{\CA,t,N}$: either the sequence of states is rejecting, or mode $\fb$ is played infinitely often.
It follows that the play of $\CG^N_\CA$ satisfies \itemfont{B} and therefore is winning for $\eve$, as requested.

This concludes the proof of \cref{lem:completeness}, hence also of \cref{thm:main}.

\section{Definability of ranks in MSO}
\label{sec:definability}

We now move to a~study of definability of particular rank bounds in MSO.

With some analogy to the cardinality quantifier by B\'ar\'any et al.~\cite{barany_expressing_trees},
one can propose a~quantifier $\exists^{\leq \eta} X.\,\phi(\vec{Y},X)$, expressing that there exists a~well\=/founded set $X$ of rank at most~$\eta$ such that $\phi(\vec{Y},X)$ holds.
Note that this can be equivalently rewritten using a~predicate $\rank(X)\leq \eta$.
Indeed, these two notions are interdefinable: $\exists^{\leq \eta} X.\,\phi(\vec{Y},X)$ can be written as $\exists X.\,\rank(X)\leq \eta\land\phi(\vec{Y},X)$,
while $\rank(X)\leq \eta$ can be written as $\exists^{\leq \eta} Z.\,X=Z$.
Below we show that the predicate $\rank(X)\leq \eta$, and consequently the respective quantifier, cannot be expressed in MSO except for the basic case of natural numbers (or $\omega_1$).

Definability of this predicate in MSO boils down to checking if the following language is regular:
\[L_{\leq\eta}\eqdef \big\{x\in\trees_{\{0,1\}}\mid \rank(x)\leq \eta\big\}.\]
Analogously, we can consider a~predicate $\rankR(X)\leq \eta$, and a~corresponding language
\[L^{\sR}_{\leq\eta}\eqdef \big\{x\in\trees_{\{0,1\}}\mid \rankR(x)\leq \eta\big\}.\]
Clearly, both $\rank(x)<\omega_1$ and $\rankR(x)<\omega_1$ hold for every well\=/founded tree, thus it remains to consider $\eta<\omega_1$.

We first show that these languages are regular for $\eta<\omega$.

\begin{fact}\label{ft:rank-definable}
	For every $\ell<\omega$ the languages $L_{\leq\ell}$ and $L^{\sR}_{\leq\ell}$ are regular.
\end{fact}

\begin{proof}
	It is easy to write an~MSO formula which given two labellings $x_i,x_{i+1}\in\trees_{\{0,1\}}$ of the infinite binary tree checks whether $x_{i+1}=\der(x_i)$.
	To express $L_{\leq\ell}$ in MSO we quantify existentially over $\ell+1$ sets $x_0,x_1,\dots,x_\ell$ and say that $x_0$ equals the input tree,
	every $x_{i+1}$ is a~derivative of $x_i$, and $x_\ell$ is empty.
	
	For $L^{\sR}_{\leq\ell}$ we proceed analogously, except that now we check whether $x_{i+1}=\derF(x_i)$.
\end{proof}

Going beyond $\omega$, the languages stop being regular.
First, a~simple pumping argument shows the following.

\begin{lem}\label{lem:rank-not-definable}
	For all ordinals $\eta<\omega_1$ and all $\ell<\omega$ the language $L_{\leq \eta+\omega+\ell}$ is not regular.
\end{lem}

\begin{proof}
	Let $x_\eta$ be a~tree of rank precisely $\eta$.
	Assume for the sake of contradiction that the language $L_{\leq \eta+\omega+\ell}$ is regular and recognised by an~automaton $\CA$ with a~set of states $Q$.
	Denote $n=|Q|+1$.

	\begin{figure}
	\begin{center}
	\tikzstyle{tnode}=[scale=0.5, anchor=mid]
	\tikzstyle{brace}=[draw, decoration={brace, mirror, amplitude=3mm}, decorate]
	\begin{tikzpicture}
	\eval{\ddx}{0.5}
	\eval{\ddy}{0.3}
	\foreach \ck in {0,...,3} {
		\node[tnode] (r\ck) at (0-\ck*\ddx,0-\ck*\ddy) {$1$};
	}
	\foreach \ck in {0,...,2} {
		\evalInt{\ckp}{\ck+1}
		\draw (r\ck) -- (r\ckp);
	}
	\coordinate(l) at (r3);
	
	\foreach \ci in {1,...,7} {
		\node[tnode] (ri\ci) at ($(l)+(0+\ci*\ddx,0-\ci*\ddy)$) {$0$};
	
		\foreach \ck in {1,...,4} {
			\node[tnode] (rik\ck) at ($(ri\ci)+(0-\ck*\ddx,0-\ck*\ddy)$) {$1$};
		}
		\coordinate (rik5) at ($(ri\ci)+(0-5*\ddx,0-5*\ddy)$);
		\draw (ri\ci) -- (rik1);
		\foreach \ck in {1,...,4} {
			\evalInt{\ckp}{\ck+1}
			\draw (rik\ck) -- (rik\ckp);
		}
		\coordinate (vc) at ($(rik5)+(0+0.0*\ddx,0-1.2*\ddy)$);
		\coordinate (vl) at ($(rik5)+(0-0.4*\ddx,0-1.5*\ddy)$);
		\coordinate (vr) at ($(rik5)+(0+0.4*\ddx,0-1.5*\ddy)$);
		\draw (rik5) -- (vl);
		\draw (rik5) -- (vr);
		\node[tnode,scale=1.2] at (vc) {$x_\eta$};
	}
	\coordinate (ri8) at ($(l)+(0+8*\ddx,0-8*\ddy)$);
	\draw (r3) -- (ri1);
	\foreach \ci in {1,...,6} {
		\evalInt{\cip}{\ci+1}
		\draw (ri\ci) -- (ri\cip);
	}
	\draw (ri7) edge[dotted] (ri8);
	
	\draw (+0.0*\ddx, +0.5*\ddy*1.44) edge[brace] node[yshift=5mm, xshift=-2mm, scale=0.8] {$\ell{+}1$} ++(-3.4*\ddx,-3.4*\ddy);
	
	\draw ($(l)+(+0.0*\ddx, -0.6*\ddy*1.44)$) edge[brace] node[yshift=5mm, xshift=-2mm, scale=0.8] {$n$} ++(-3.4*\ddx,-3.4*\ddy);
	\end{tikzpicture}
	\end{center}
	\caption{An~illustration of the tree $x$ used in the proof of \cref{lem:rank-not-definable}}
	\label{fig:tree-for-pumping}
	\end{figure}

	Consider a~tree $x$ (see \Cref{fig:tree-for-pumping} for an~illustration of this tree) such that
	\begin{itemize}
	\item for every $i\leq \ell$ we have $x\big(\dL^i\big)=1$---there are exactly $\ell{+}1$ nodes labelled by~$1$ on the left\=/most branch of $x$;
	\item for every $i>0$ and $j<n$ we have $x\big(\dL^{\ell}\dR^i\dL^{j+1}\big)=1$---there are exactly $n$ nodes labelled by~$1$ on the left\=/most branches of $x$ which begin in nodes of the form $\dL^{\ell}\dR^i\dL$;
	\item for every $i>0$ the subtree $x\restr_{\dL^{\ell}\dR^i\dL^{n+1}}$ equals $x_\eta$---at the end of these sequences of $n$ labels~$1$ we see the subtree $x_\eta$;
	\item all the remaining nodes of $x$ are labelled by $0$.
	\end{itemize}
	\noindent 
	It is easy to see that $\rank(x)=\eta+n+\ell+1<\eta+\omega\leq\eta+\omega+\ell$ so this tree belongs to $L_{\leq \eta+\omega+\ell}$.
	Let $\rho$ be an~accepting run of $\CA$ on $x$.
	By the definition of $n$, this run has a~repetition of a~state on every sequence of $n$ consecutive nodes: for each $i>0$ there exist $0\leq j_i<j'_i<n$ such that
	\[\rho\big(\dL^{\ell}\dR^i\dL^{j_i+1}\big)=\rho\big(\dL^{\ell}\dR^i\dL^{j'_i+1}\big).\]
	By repeating this part of the run exactly $i$ times we obtain an~accepting run $\rho'$ of $\CA$ on a~well\=/founded tree $x'$.
	However, $\rank(x')=\eta+\omega+\ell+1>\eta+\omega+\ell$, a~contradiction.
\end{proof}

\begin{cor}\label{cor:not-expressible}
	For an~ordinal $\eta<\omega_1$ the language $L_{\leq\eta}$ is regular if and only if $\eta<\omega$.
	The same holds for $L^{\sR}_{\leq\eta}$.
\end{cor}

\begin{proof}
	Definability of ranks and layered ordinals for $\eta<\omega$ follows from \cref{ft:rank-definable}.

	Now assume that the language $L_{\leq\eta}$ is regular for some $\eta<\omega_1$.
	Consider the relation $\Gamma=\big\{(x,x)\mid x\in L_{\leq\eta}\big\}$.
	Then $\limitR(\Gamma)=\eta$ and \cref{thm:main} implies that $\eta <\omega^2$.
	Thus, either $\eta<\omega$ and the thesis holds, or $\eta=\eta'+\omega+l$ for some ordinal $\eta'$ and $l\in\N$.
	In the latter case \cref{lem:rank-not-definable} contradicts regularity of $L_{\leq\eta}$.

	Next, assuming that $L^{\sR}_{\leq\eta}$ is regular for some $\eta<\omega_1$ consider the relation $\Gamma=\big\{(x,x)\mid x\in L^{\sR}_{\leq\eta}\big\}$.
	Then $\limitF(\Gamma)=\eta$ and \cref{pro:main-layers} implies that $\eta <\omega$.
\end{proof}

The above \lcnamecref{cor:not-expressible} does not exclude the
possibility that a~higher ordinal $\eta$ may be a~supremum of a~regular
subset of $L_{\leq\eta}$.

\begin{lem}\label{lem:construct-ordinals}
	For each pair of natural numbers $k$, $\ell$ there exists a~regular language $L\subseteq\WF$ such that $\sup_{x\in L} \rank(x)=\omega\cdot k+\ell$.
\end{lem}

\begin{proof}
	It is easy to see that given a~language $L$, one can define a~new language $L'$ such that $\sup_{x\in L'}\rank(x)=1+\sup_{x\in L}\rank(x)$.
	The language $L'$ explicitly requires an~additional $1$ in the root of the considered tree~$x$ and checks if $x\restr_\dL\in L$ and if $x\restr_\dR$ is empty (i.e.,~has all nodes labelled by $0$).

	Thus, it is enough to show, for each $k\in\N$, a~language $L_k$ such that $\sup_{x\in L_k}\rank(x)=\omega\cdot k$.
	Let $L_0$ contain only the tree $x_0$ with all nodes labelled by $0$ (i.e., the unique tree of rank~$0$).
	Let $L_{k+1}$ contain a~tree~$x\in \WF$ if for every $i\in\N$ there exists $\ell_i\in\N$ such that
	\begin{itemize}
	\item $x(\dR^i)=0$,
	\item $x(\dR^i\dL^{j+1})=1$ for $j<\ell_i$,
	\item $x\restr_{\dR^i\dL^{\ell_i+1}}$ belongs to $L_k$,
	\item all the remaining nodes of $x$ are labelled by $0$.
	\end{itemize}
	It is easy to see that if $L_k$ is regular then so is $L_{k+1}$.
	Moreover, if every tree $x\in L_k$ has rank at most $\omega\cdot k$ then directly by the definition, every tree $x\in L_{k+1}$ has rank at most $\omega\cdot (k{+}1)$.
	Additionally, if we choose the subtrees $x\restr_{\dR^i\dL^{\ell_i+1}}$ in such a~way that their rank is precisely $\omega\cdot k$ and make the values $\ell_i$ tend to infinity with $i$,
	the rank of the obtained tree $x$ equals $\omega\cdot (k{+}1)$.
	Therefore, $\sup_{x\in L_{k+1}}\rank(x)=\omega\cdot (k{+}1)$.
\end{proof}

The above results show that $\rankR$ behaves in a~more robust way than $\rank$, when talking about definability in MSO.
While the dichotomy theorem works for both ranks, showing that each of them becomes undefinable above some threshold ($\omega$ and $\omega^2$ respectively)
here we see that it is possible to define all values of $\rankR$ up to $\omega$, while it is not possible for $\rank$ and $\omega^2$.

\begin{rem}\label{rem:wmsou}
	Clearly no countable formalism can define all the languages $L_{\leq\eta}$ for $\eta<\omega_1$.
	However, certain formalisms can go beyond MSO.
	For instance, the logic WMSO+U (for which the satisfiability problem is known to be decidable~\cite{bojanczyk_wmso_u_p}) is capable of defining the language $L_{\leq \omega}$:
	a~tree $x$ has rank at most $\omega$ if below every node~$u$ labelled by~$1$ there is a~bound~$K$ on the number of nodes labelled by~$1$ that can appear on branches of $x$ passing through $u$.
	This construction can be iterated to define the languages $L_{\leq \omega\cdot k+\ell}$ and possibly even beyond that.
\end{rem}

\section{Closure ordinals}
\label{sec:closure_ordinals}

In this section we show how a~negative answer to Czarnecki's question on closure ordinals can be derived from the present result.
We use the standard syntax and semantics of modal $\mu$\=/calculus~\cite{kozen-mu-calc,niwinski_rudiments,Demri2016,julian-igor-handook}, with its formulae constructed using the following grammar:
\[F ::= a \mid X \mid \mu X.\, F \mid \nu X.\, F \mid F_1\lor F_2 \mid F_1\land F_2 \mid \Diamond F \mid \Box F,\]
where $a\in A$ is a~letter from a~fixed alphabet, $X$ is a~variable from some fixed set of variables,
$\mu$ and $\nu$ are the least and the greatest \emph{fixed\=/point operators}, and $\Diamond$ and $\Box$ are the standard modalities (``exists a~successor'' and ``for all successors'').
For technical convenience we make an~assumption that, in each point of our model, exactly one proposition (letter in $A$) is satisfied.
For the sake of readability we often identify a~formula with its semantics, that is, we read a~closed formula as a~subset of the domain,
and a~formula with $k$ free variables as a~$k$\=/ary function over its powerset.

Additionally, we allow a~single vectorial $\mu$ operator on the very outside of the considered formula.
Namely, for $k\geq 1$ we consider a~tuple of $k$ variables $\vec{X}=(X_1,\dots,X_k)$ and a~tuple of $k$ formulae $\vec{F}(\vec{X})=(F_1(\vec{X}),\dots,F_k(\vec{X}))$,
each of them having free variables in $X_1,\dots,X_k$.
Then $\mu\vec{X}.\,\vec{F}(\vec{X})$ denotes the tuple of sets being the least fixed point of the function $\vec{X}\mapsto\vec{F}(\vec{X})$.

Given an~ordinal number $\eta$, we use the standard notation $\mu^\eta \vec{X}.\,\vec{F}(\vec{X})$ for the $\eta$\=/approximation of the fixed point in a~given model
(which amounts to the $\eta$'s iteration $\vec{F}^{\eta}(\emptyset,\dots,\emptyset)$).

The following definition introduces closure ordinals for $\mu$\=/calculus formulae, as considered in \cref{con:czarnecki}.

\begin{defi}
	Given a~model $\tau$, we define the \emph{closure ordinal of $\mu \vec{X}.\, \vec{F}(\vec{X})$ in $\tau$} as the least ordinal $\eta$ such that $\mu \vec{X}.\,\vec{F}(\vec{X}) = \mu^\eta \vec{X}.\,\vec{F}(\vec{X})$.
	The \emph{closure ordinal of $\mu \vec{X}.\,\vec{F}(\vec{X})$} is the supremum of these ordinals over all models $\tau$ (or $\infty$ if the supremum does not exist).
\end{defi}

We aim at providing an~(alternative to Afshari, Barlucchi, and Leigh~\cite{afshari_limit_fics}) proof of the following \lcnamecref{thm:closure-mu}.

\begin{thm}\label{thm:closure-mu}
	Let $\vec{F}(\vec{X})=(F_1(\vec{X}),\dots,F_k(\vec{X}))$ with $\vec{X}=(X_1,\dots,X_k)$ be a~tuple of $\mu$\=/calculus formulae in which the variables $X_1,\dots,X_k$ do not occur in the scope of any fixed\=/point operator.
	Then, the closure ordinal of $\mu \vec{X}.\,\vec{F}(\vec{X})$ is either strictly smaller than $\omega^2$, or at least $\omega_1$, and it can be effectively decided which of the cases holds.
	In the former case, it is possible to compute a~number $N\in\N$ such that the closure ordinal is bounded by $\omega\cdot N$.
\end{thm}
\noindent 
Note that we allow arbitrary closed formulae of $\mu$\=/calculus to be nested in $\vec{F}$;
however, we do not cover the whole $\mu$\=/calculus, because of the restriction on occurrences of the variables of $\vec{X}$.
This stays in line with the fragment considered by Afshari, Barlucchi,
and Leigh~\cite{afshari_limit_fics} (as explained in \cref{sec:intro}),
but we additionally provide a~decision procedure that makes the
dichotomy effective.

\begin{rem}
	One could consider a~richer syntax of $\mu$\=/calculus, with vectorial fixed\=/point operators allowed also inside $\vec{F}$, together with some construction for extracting particular coordinates out of a~tuple.
	This, however, does not change anything: formulae of such vectorial $\mu$\=/calculus are easily equivalent to formulae of standard $\mu$\=/calculus,
	where we appropriately nest fixed\=/point operators for particular variables of a~tuple. This equivalence is often referred to as Bekić Principle (see, e.g.,~\cite{Bekic1984}).

	On the other hand, by considering $\mu \vec{X}.\,\vec{F}(\vec{X})$ for a~tuple of variables and formulae (instead of $\mu X.\,F(X)$ for a~single variable and a~single formula)
	we make \cref{thm:closure-mu} formally stronger.
	Namely, the closure ordinal counts the number of iterations for the whole tuple, not just for a~single variable.
	Moreover, while converting $\mu \vec{X}.\,\vec{F}(\vec{X})$ into a~formula with nested $\mu$ operators for particular variables,
	one would usually obtain a~formula with $X_1$ occurring in the scope of the $\mu$~operator for $X_2$, making it impossible to use the \lcnamecref{thm:closure-mu}.
\end{rem}

Towards a~proof of \cref{thm:closure-mu}, as a~first step, we eliminate from $F_1,\dots,F_k$ all occurrences of $X_1,\dots,X_k$ that are not in the scope of any modal operator.
While it is known~\cite{kozen-mu-calc} that such an~elimination is possible (leading to so\=/called guarded formulae), here we need to do it carefully,
so that we can bound how the closure ordinal changes after the transformation.
This change vanishes when the closure ordinal is a~limit ordinal, which is the case we are interested in.

\begin{lem}\label{lem:guarded}
	Let $\vec{F}(\vec{X})=(F_1(\vec{X}),\dots,F_k(\vec{X}))$ with $\vec{X}=(X_1,\dots,X_k)$ be a~tuple of $\mu$\=/calculus formulae in which the variables $X_1,\dots,X_k$ do not occur in the scope of any fixed\=/point operator,
	such that the closure ordinal of $\mu\vec{X}.\,\vec{F}(\vec{X})$ is bounded by a~limit ordinal $\eta$.
	Then there exists a~tuple of formulae $\vec{H}(\vec{X})=(H_1(\vec{X}),\dots,H_k(\vec{X}))$ such that
	\[
		\mu\vec{X}.\,\vec{F}(\vec{X}) = \mu\vec{X}.\,\vec{H}(\vec{X}),
	\]
	the closure ordinal of $\mu\vec{X}.\,\vec{H}(\vec{X})$ is also bounded by $\eta$,
	the variables $X_1,\dots,X_k$ do not occur in $H_1,\dots,H_k$ in the scope of any fixed\=/point operator,
	and moreover all their occurrences are in the scope of a~modal operator.
\end{lem}

\begin{proof}
	The assumption that the variables $X_1,\dots,X_k$ do not occur in the scope of any fixed\=/point operator in the constructed formulae is maintained throughout the proof without explicitly mentioning it.

	We prove that it is possible to perform two kinds of more basic transformations, where $i\in\{1,\dots,k\}$:
	\begin{enumerate}
	\item	Suppose that all occurrences of $X_1,\dots,X_{i-1}$ in $F_1,\dots,F_k$ are already in the scope of a~modal operator.
		Then we can ensure that the same holds for $\vec{H}$, and additionally also all occurrences of $X_i$ in $H_i$ are in the scope of a~modal operator.
	\item	Suppose that all occurrences of $X_1,\dots,X_{i-1}$ in $F_1,\dots,F_k$ and all occurrences of $X_i$ in $F_i$ are already in the scope of a~modal operator.
		Then we can ensure that all occurrences of $X_1,\dots,X_i$ in $H_1,\dots,H_k$ are in the scope of a~modal operator.
	\end{enumerate}
	\noindent 
	For the first transformation, fix some $i\in\{1,\dots,k\}$.
	Using distributivity laws we transform $F_i(X)$ to a~disjunctive normal form; the result can be written as
	\begin{align*}
		\big(X_i\land G_1(\vec{X})\big)\lor\dots\lor\big(X_i\land G_n(\vec{X})\big)\lor H(\vec{X}),
	\end{align*}
	where $X_i$ occurs in all $G_j$ and in $H$ only in the scope of modal operators
	(in general $X_i$ could also occur in the scope of fixed\=/point operators, but assumptions of the lemma forbid this).
	Using distributivity laws again, we can group $G_1,\dots,G_n$ into their disjunction $G$, obtaining that $F_i(\vec{X})$ is equivalent to
	\begin{align*}
		\big(X_i\land G(\vec{X})\big)\lor H(\vec{X}).
	\end{align*}
	We then take
	\begin{align*}
		\vec{H}(\vec{X})=(F_1(\vec{X}),\dots,F_{i-1}(\vec{X}),H(\vec{X}),F_{i+1}(\vec{X}),\dots,F_n(\vec{X})).
	\end{align*}
	
	We now fix a~model, and prove that $\mu^\gamma\vec{X}.\,\vec{F}(\vec{X})=\mu^\gamma\vec{X}.\,\vec{H}(\vec{X})$ for every ordinal $\gamma$,
	which implies that $\mu\vec{X}.\,\vec{F}(\vec{X})$ and $\mu\vec{X}.\,\vec{H}(\vec{X})$ have precisely the same closure ordinal, and the same value.
	This is done by transfinite induction on $\gamma$.
	For limit ordinals this follows immediately from the induction hypothesis.
	For a~successor ordinal $\gamma+1$ denote $\vec{Y}=\mu^\gamma\vec{X}.\,\vec{F}(\vec{X})=\mu^\gamma\vec{X}.\,\vec{H}(\vec{X})$ (with equality by the induction hypothesis).
	Now, $\mu^{\gamma+1}\vec{X}.\,\vec{F}(\vec{X})=\vec{F}(\vec{Y})$ and $\mu^{\gamma+1}\vec{X}.\,\vec{H}(\vec{X})=\vec{H}(\vec{Y})$.
	On coordinates other than the $i$\=/th we apply literally the same functions to $\vec{Y}$, so those coordinates of $\vec{F}(\vec{Y})$ and $\vec{H}(\vec{Y})$ are equal.
	Note that $\vec{Y}=(Y_1,\dots,Y_k)$, being $\mu^\gamma\vec{X}.\,\vec{H}(\vec{X})$, is a~pre\=/fixed point of $\vec{H}$, that is, satisfies $\vec{Y}\subseteq\vec{H}(\vec{Y})$ (coordinatewise inclusion);
	in particular $Y_i\subseteq H(\vec{Y})$.
	We thus obtain equality also on the $i$\=/th coordinate:
	\begin{align*}
		F_i(\vec{Y})=\big(Y_i\land G(\vec{Y})\big)\lor H(\vec{Y})=H(\vec{Y}).
	\end{align*}
	
	We now come to the second transformation.
	Again, $i\in\{1,\dots,k\}$ is fixed.
	This time we do not change the $i$\=/th formula, taking $H_i=F_i$.
	For $j\neq i$ we obtain $H_j$ from $F_j$ by substituting $F_i(\vec{X})$ for every occurrence of $X_i$.
	Recall that all occurrences of $X_i$ in $F_i$ are in the scope of a~modal operator, so after the above substitution the same holds also for the resulting formula $H_j$.

	We now fix a~model, and prove that $\mu^\gamma\vec{X}.\,\vec{F}(\vec{X})\subseteq\mu^\gamma\vec{X}.\,\vec{H}(\vec{X})$ for every ordinal~$\gamma$, by transfinite induction on $\gamma$.
	For limit ordinals this follows immediately from the induction hypothesis.
	For a~successor ordinal $\gamma+1$ denote $\vec{Y}=\mu^\gamma\vec{X}.\,\vec{F}(\vec{X})$.
	By definition $\mu^{\gamma+1}\vec{X}.\,\vec{F}(\vec{X})=\vec{F}(\vec{Y})$, while the induction hypothesis $\vec{Y}\subseteq\mu^\gamma\vec{X}.\,\vec{H}(\vec{X})$ by monotonicity of $\vec{H}$ implies that
	$\vec{H}(\vec{Y})\subseteq\mu^{\gamma+1}\vec{X}.\,\vec{H}(\vec{X})$; it is thus enough to prove that $\vec{F}(\vec{Y})\subseteq\vec{H}(\vec{Y})$.
	On the $i$\=/th coordinate this is automatic: the same function is applied to $\vec{Y}=(Y_1,\dots,Y_k)$.
	For every other coordinate $j$ we note that $H_j(\vec{Y})=F_j(Y_1,\dots,Y_{i-1},F_i(\vec{Y}),Y_{i+1},\dots,Y_k)$.
	We then use the fact that $\vec{Y}$ is a~pre\=/fixed point of $\vec{F}$, meaning that $\vec{Y}\subseteq\vec{F}(\vec{Y})$:
	we have $Y_i\subseteq F_i(\vec{Y})$, which by monotonicity of $F_j$ implies the thesis
	\begin{align*}
		F_j(Y_1,\dots,Y_k)\subseteq F_j(Y_1,\dots,Y_{i-1},F_i(\vec{Y}),Y_{i+1},\dots,Y_k)=H_j(Y_1,\dots,Y_k).
	\end{align*}

	Likewise, we prove that $\mu^\gamma\vec{X}.\,\vec{H}(\vec{X})\subseteq\mu^{2\gamma}\vec{X}.\,\vec{F}(\vec{X})$ for every ordinal $\gamma$; again, by transfinite induction on $\gamma$,
	and again only the case of a~successor ordinal $\gamma+1$ is interesting.
	Denote $\vec{Y}=\mu^{2\gamma}\vec{X}.\,\vec{F}(\vec{X})$.
	Because $\mu^{\gamma+1}\vec{X}.\,\vec{H}(\vec{X})=H(\mu^\gamma\vec{X}.\,\vec{H}(\vec{X}))\subseteq \vec{H}(\vec{Y})$ (by the induction hypothesis and monotonicity of $H$)
	and $\mu^{2(\gamma+1)}\vec{X}.\,\vec{F}(\vec{X})=\vec{F}(\vec{F}(\vec{Y}))$ (by definition), it is enough to prove that $\vec{H}(\vec{Y})\subseteq\vec{F}(\vec{F}(\vec{Y}))$.
	We again use the fact that $\vec{Y}$ is a~pre\=/fixed point of $\vec{F}$: we have $\vec{Y}\subseteq\vec{F}(\vec{Y})$, that is, $Y_j\subseteq F_j(\vec{Y})$ for all $j$.
	On the $i$\=/th coordinate we simply have $H_i(\vec{Y})=F_i(\vec{Y})\subseteq F_i(\vec{F}(\vec{Y}))$,
	and for $j\neq i$ we have
	\begin{align*}
		H_j(\vec{Y})=F_j(Y_1,\dots,Y_{i-1},F_i(\vec{Y}),Y_{i+1},\dots,Y_k)\subseteq F_j(F_1(\vec{Y}),\dots,F_k(\vec{Y}))=F_j(\vec{F}(\vec{Y})),
	\end{align*}
	which proves the thesis.
	
	For limit ordinals $\gamma$, in particular for $\gamma=\eta$, we have $2\cdot\gamma=\gamma$, so the above gives us the equality $\mu^\gamma\vec{X}.\,\vec{F}(\vec{X})=\mu^\gamma\vec{X}.\,\vec{H}(\vec{X})$.
	It follows that the closure ordinals of the two fixed\=/point formulae are bounded by the same limit ordinals (although they may differ by a~finite number);
	moreover $\mu\vec{X}.\,\vec{F}(\vec{X})=\mu\vec{X}.\,\vec{H}(\vec{X})$.
\end{proof}

Next, using rather standard techniques we obtain the following two \lcnamecrefs{lem:mu-tree1}.
In these \lcnamecrefs{lem:mu-tree1}, when $\tau$ is a~tree structure (i.e.,~a~node\=/labelled directed graph with a~distinguished root node from which every other node of $\tau$ is reachable in exactly one way),
we write $\rank(Z)$ for the rank of a~set $Z\subseteq\dom(\tau)$, defined in the same way as for a~set $Z\subseteq\{\dL,\dR\}^\ast$ of nodes of the binary tree.
For a~tuple of sets $\vec{Z}=(Z_1,\dots,Z_k)$ by $\pi_i(\vec{Z})$ we denote its $i$\=/th coordinate $Z_i$.

\begin{lem}\label{lem:mu-tree1}
	Let $\vec{F}(\vec{X})=(F_1(\vec{X}),\dots,F_k(\vec{X}))$ with $\vec{X}=(X_1,\dots,X_k)$ be a~tuple of formulae as in \cref{thm:closure-mu}, with all occurrences of $X_1,\dots,X_k$ being in the scope of a~modal operator.
	The following two conditions are equivalent for every countable ordinal $\eta$:
	\begin{itemize}
	\item	the closure ordinal of $\mu \vec{X}.\,\vec{F}(\vec{X})$ is bounded by $\eta$;
	\item	for every model $\tau$ that is a~countable tree with root $r$ we have $r\in\pi_i(\mu^\eta\vec{X}.\,\vec{F}(\vec{X}))$ for all $i\in\{1,\dots,k\}$ such that $r\in\pi_i(\mu \vec{X}.\,\vec{F}(\vec{X}))$.
	\end{itemize}
\end{lem}

\begin{proof}
	Implication ``$\Rightarrow$'' is trivial: by the definition of a~closure ordinal we have $\mu\vec{X}.\,\vec{F}(\vec{X})=\mu^\eta\vec{X}.\,\vec{F}(\vec{X})$ in every model $\tau$.
	
	For the implication ``$\Leftarrow$'' suppose that the first item is false: the closure ordinal of $\mu\vec{X}.\,\vec{F}(\vec{X})$ is greater than $\eta$.
	Then, by definition of a~supremum, there exists a~particular model $\tau$ such that the closure ordinal of $\mu\vec{X}.\,\vec{F}(\vec{X})$ in $\tau$ is greater than $\eta$,
	that is, $\mu\vec{X}.\,\vec{F}(\vec{X})\neq\mu^\eta\vec{X}.\,\vec{F}(\vec{X})$.
	This inequality implies that there exists a~coordinate $i\in\{1,\dots,k\}$ and a~node $r\in\dom(\tau)$ which belongs to $\pi_i(\mu\vec{X}.\,\vec{F}(\vec{X}))$, but not to $\pi_i(\mu^\eta\vec{X}.\,\vec{F}(\vec{X}))$.
	Fix $\tau$, $i$, and $r$, and consider the tree $\tau'$ obtained by unfolding $\tau$ in $r$.
	Because $\mu$\=/calculus is invariant under bisimulation
	(cf.~\cite[Section~26.3.1]{julian-igor-handook}), the root $r'$ of $\tau'$ belongs to $\pi_i(\mu\vec{X}.\,\vec{F}(\vec{X}))$.
	Simultaneously, it is easy to prove by transfinite induction on $\gamma$ that a~node of the tree~$\tau'$ belongs to $\pi_j(\mu^\gamma\vec{X}.\,\vec{F}(\vec{X}))$, for some $j$,
	if and only if the corresponding node of the original structure $\tau$ belongs to $\pi_j(\mu^\gamma\vec{X}.\,\vec{F}(\vec{X}))$.
	In particular this holds for $\gamma=\eta$ and for the root $r'$: the root $r'$ does not belong to $\pi_i(\mu^\eta\vec{X}.\,\vec{F}(\vec{X}))$.
	
	To make the model countable, we need to recall the semantics
	of $\mu$\=/calculus in terms of games (see~\cite[Section 26.2.3]{julian-igor-handook}):
	in order to evaluate formulae $\vec{F}(\vec{X})$ in a~given model~$\tau'$, for a~given valuation $X_1,\dots,X_k\subseteq\dom(\tau')$ of the variables of $\vec{X}$,
	one defines an \emph{evaluation game} (called \emph{verification game} in \cite{julian-igor-handook}), so that
	the satisfaction of the formula is determined by the winner of that game.
	This is a~game with positions of the form $(x,G)$, where $x\in\dom(\tau')$ and where $G$ is a~subformula of some $F_j$
	(assuming that every bound variable of $\vec{F}$ has a~different name).
	An~important detail is that there may be a~move from $(x,G)$ to $(x',G')$ only when either $x'=x$ or $x'$ is a~successor of $x$.
	For every ordinal $\gamma\leq\eta$ we consider such a~game for $\vec{F}(\vec{X})$ in the setting where $\vec{X}$ is valuated to $\mu^\gamma\vec{X}.\,\vec{F}(\vec{X})$.
	We also consider one more game for evaluating the closed formula $\mu\vec{X}.\,\vec{F}(\vec{X})$.
	The evaluation games are games with parity winning condition, so they are positionally determined.
	Thus, in each of the countably many games fix a~pair of positional strategies (one per player), which for every node are winning for the player who wins from this node.
	Having them, we say that a~successor $x'$ of a~node $x$ is \emph{important}
	if from some position $(x,G)$ in some of the considered games the positional strategy requests a~move to a~position $(x',G')$ involving this successor.
	For each $x$ there are finitely many positions $(x,G)$, and for each of the countably many games, the positional strategy points just to a~single successor position
	(or to none, if the position does not belong to the winning player); thus every node has only countably many important successors.
	We can now prune the tree: for every node $x$ of $\tau'$ we remove all its non\=/important successors, together with the whole subtrees starting there.
	We obtain a~new tree $\tau''$, in which every node has countably many successors, hence the whole $\tau''$ is countable.
	Observe also that the root $r''$ of $\tau''$ still belongs to $\pi_i(\mu\vec{X}.\,\vec{F}(\vec{X}))$.
	Indeed, the winning strategy from the evaluation game for $\tau'$ still works in the evaluation game for $\tau''$; we have not removed any positions to which the strategy requested to move.
	Furthermore, we claim that for every $\gamma\leq\eta$ the tuple of sets $\mu^\gamma\vec{X}.\,\vec{F}(\vec{X})$ calculated in $\tau''$
	is just the restriction to $\tau''$ of the tuple $\mu^\gamma\vec{X}.\,\vec{F}(\vec{X})$ calculated in $\tau'$.
	We prove this by transfinite induction on $\gamma$.
	For limit ordinals this follows immediately from the induction hypothesis.
	For ordinals of the form $\gamma+1$ we have $\mu^{\gamma+1}\vec{X}.\,\vec{F}(\vec{X})=\vec{F}(\mu^\gamma\vec{X}.\,\vec{F}(\vec{X}))$.
	By the induction hypothesis we know that the evaluation game for $\vec{F}$ and $\tau''$,
	where $\vec{X}$ is valuated $\mu^\gamma\vec{X}.\,\vec{F}(\vec{X})$, is obtained by removing some positions from the game for $\vec{F}$ and $\tau'$
	(but, importantly, $\vec{X}$ in the remaining nodes is evaluated in an~unchanged way).
	We have not removed any positions to which our winning strategies requested to move, so the strategies are still winning;
	thus every node of $\tau''$ belongs to $\pi_j(\vec{F}(\mu^\gamma\vec{X}.\,\vec{F}(\vec{X})))$ for some $j$
	if and only if the corresponding node of $\tau'$ belongs to $\pi_j(\vec{F}(\mu^\gamma\vec{X}.\,\vec{F}(\vec{X})))$, which gives the claim for $\gamma+1$.
	We have thus obtained a~countable tree~$\tau''$ whose root $r''$ belongs to $\pi_i(\mu\vec{X}.\,\vec{F}(\vec{X}))$ but not to $\pi_i(\mu^\eta\vec{X}.\,\vec{F}(\vec{X}))$, contradicting the second item.
\end{proof}

\begin{lem}\label{lem:mu-tree2}
	Let $\vec{F}(\vec{X})=(F_1(\vec{X}),\dots,F_k(\vec{X}))$ with $\vec{X}=(X_1,\dots,X_k)$ be a~tuple of formulae as in \cref{thm:closure-mu}, with all occurrences of $X_1,\dots,X_k$ being in the scope of a~modal operator.
	Let $\tau$ be a~tree model, and $r$ its node.
	The following two conditions are equivalent for every limit ordinal $\eta$ and every $I\subseteq\{1,\dots,k\}$:
	\begin{itemize}
	\item	$r\in\pi_i(\mu^\eta\vec{X}.\,\vec{F}(\vec{X}))$ for all $i\in I$;
	\item	there exists a~tuple $\vec{Z}$ of well\=/founded sets $Z_1,\dots,Z_k\subseteq\dom(\tau)$
		such that $\vec{F}(\vec{Z})\supseteq\vec{Z}$, and $\rank(Z_1\cup\dots\cup Z_k)\leq\eta$, and $r\in Z_i$ for all $i\in I$.
	\end{itemize}
\end{lem}

\begin{proof}
	For the whole proof fix a~tree model $\tau$ and its node $r$.

	Let us start with the implication ``$\Rightarrow$''.
	For $x,y\in\dom(\tau)$ let $\dist(x,y)$ denote the number of steps (i.e.,~edges) needed to reach $y$ from $x$, or $\infty$ if $y$ is not reachable from~$x$.
	Let $D\in\N$ be a~number such that the variables $X_1,\dots,X_k$ occur in the formulae $F_1(\vec{X}),\dots,F_k(\vec{X})$ only in the scope of at most $D$ modal operators.
	Recall also that, by assumption, those variables do not occur in the scope of fixed\=/point operators.
	Thus, while evaluating $\vec{F}(\vec{X})$ at some node $x$, we can examine the sets $X_1,\dots,X_k$ only at nodes at distance at most $D$ from $x$.
	Because additionally $\vec{F}$ is monotone in $\vec{X}$, we can formulate the following property,
	which holds for every $x\in\dom(\tau)$, every $i\in\{1,\dots,k\}$, and all tuples $\vec{X}=(X_1,\dots,X_k),\vec{Y}=(Y_1,\dots,Y_k)$ of subsets of $\dom(\tau)$:
	\begin{align}
		\mbox{if }\Big(\forall y,j.\,\dist(x,y)\leq D\Rightarrow(y\in X_j\Rightarrow y\in Y_j)\Big)
		\mbox{ then }\Big(x\in F_i(\vec{X})\Rightarrow x\in F_i(\vec{Y})\Big).\tag{$\star$}\label[property]{prop-star}
	\end{align}
	Let us also denote by $\stab(x,i)$ the least ordinal $\gamma$ such that $x\in\pi_i(\mu^\gamma\vec{X}.\,\vec{F}(\vec{X}))$, with $\stab(x,i)=\infty$ if there is no such $\gamma$.
	Recall that $\pi_i(\mu^\gamma\vec{X}.\,\vec{F}(\vec{X}))$ for limit ordinals is just the union of $\pi_i(\mu^{\xi}\vec{X}.\,\vec{F}(\vec{X}))$ over all $\xi<\gamma$.
	Thus $\stab(x,i)$ is necessarily a~successor ordinal (or~$\infty$).
	We then define each $Z_i$ to be the set of those nodes $x$ for which there exists a~sequence $x_0,x_1,\dots,x_n$ of nodes and a~sequence $i_0,i_1,\dots,i_n$ of indices in $\{1,\dots,k\}$ such that
	\begin{itemize}
	\item	$x_0=r$,
	\item	$x_n=x$, $i_n=i$,
	\item	$\dist(x_{a-1},x_a)\leq D$ for all $a\in\{1,\dots,n\}$, and
	\item	$\eta\geq\stab(x_0,i_0)>\stab(x_1,i_1)>\dots>\stab(x_n,i_n)$.
	\end{itemize}
	Suppose now that $r\in\pi_i(\mu^\eta\vec{X}.\,\vec{F}(\vec{X}))$ for some limit ordinal $\eta$ and some $i\in I$.
	This can be restated as $\stab(r,i)\leq\eta$.
	Taking $x_0=r$ and $i_0=i$ for $n=0$ we immediately see that $r\in Z_i$.
	It remains to prove that $\vec{F}(\vec{Z})\supseteq\vec{Z}$ and that $\rank(Z_1\cup\dots\cup Z_k)\leq\eta$, which is obtained in the next two claims.

	\begin{clm}
		$\vec{F}(\vec{Z})\supseteq\vec{Z}$.
	\end{clm}

	\begin{proof}
		Take an~index $i\in\{1,\dots,k\}$ and a~node $x\in Z_i$, and fix sequences $x_0,x_1,\dots,x_n$ and $i_0,i_1,\dots,i_n$ as in the definition of $\vec{Z}$, ending in $x$ and $i$.
		Denote the successor ordinal $\stab(x,i)$ as $\gamma+1$.
		We then have $x\in\pi_i(\mu^{\gamma+1}\vec{X}.\,\vec{F}(\vec{X}))=F_i(\mu^\gamma\vec{X}.\,\vec{F}(\vec{X}))$.
		We now want to deduce that $x\in F_i(\vec{Z})$ from Property~($\star$) applied to the tuples $\mu^\gamma\vec{X}.\,\vec{F}(\vec{X})$ and $\vec{Z}$.
		To this end, consider an~index $j\in\{1,\dots,k\}$ and a~node $y$ such that $\dist(x,y)\leq D$ and $y\in\pi_j(\mu^\gamma\vec{X}.\vec{F}(\vec{X}))$;
		we need to see that $y\in Z_j$.
		The definition of $\stab(\cdot)$ gives us $\stab(y,j)\leq\gamma$, that is, $\stab(y,j)<\stab(x,i)$.
		It follows that the sequences $x_0,x_1,\dots,x_n,y$ and $i_0,i_1,\dots,i_n,j$ witness that $y\in Z_j$, as needed.
	\end{proof}
	
	Denote $U=Z_1\cup\dots\cup Z_k$.
	Below, by $Y\restr_x$ we denote the set of those elements of a~set~$Y$ that are reachable from a~node $x$.
	
	\begin{clm}\label{clm:2}
		Let $\gamma$ be an~ordinal and let $x\in\dom(\tau)$ be such that $\stab(y,i)\leq\gamma$ for all $i\in\{1,\dots,k\}$ and $y\in Z_i\restr_x$.
		Then $\rank(U\restr_x)\leq D\cdot(\gamma+1)$.
	\end{clm}
		
	Before proving the claim, let us see that, once it is used for $x=r$ it implies that $\rank(U)\leq\eta$.
	Indeed, if there are no $i$ for which $\stab(r,i)\leq\eta$ (which may happen if $I=\emptyset$), then all the sets $Z_i$ are empty, and we simply have $\rank(U)=0$.
	Otherwise, let $\gamma$ be the maximum of those $\stab(r,i)$ that are bounded by~$\eta$ (i.e., such that $r\in Z_i$).
	Recall that $\stab(r,i)$, hence also $\gamma$, are successor ordinals; on the other hand $\eta$ is a~limit ordinal, so $\gamma<\eta$.
	Simultaneously, for all $j\in\{1,\dots,k\}$ and $y\in Z_j$ we have $\stab(y,j)\leq\stab(r,i)\leq\eta$ for some $i$ (by the definition of $Z_j$), so $\stab(y,j)\leq\gamma$.
	Thus, the above claim yields that $\rank(U)\leq D\cdot(\gamma+1)\leq D\cdot\eta=\eta$, where the equality holds because $\eta$ is a~limit ordinal.
	
	\begin{proofof}{\cref{clm:2}.}
		The proof is by induction on $\gamma$.
		Fix a~node $x$ as in the assumptions of the claim.
		First, for every node $y$ such that $\dist(x,y)=D$ we prove that $\rank(U\restr_y)\leq D\cdot\gamma$.
		To this end, fix also such a~node $y$.
		If $U\restr_y=\emptyset$, then this is clear; assume thus that $U\restr_y\neq\emptyset$.
		Let $P_{x,y}$ contain the nodes on the path from $x$ to $y$, including $x$, but excluding $y$ (we have $|P_{x,y}|=D$).
		For every $i\in\{1,\dots,k\}$ and every $z\in Z_i\restr_y$ the definition of $Z_i$ gives us a~sequence of nodes on the path from $r$ to $z$, with consecutive nodes at distance at most $D$.
		One of these nodes, say $x'$, has to be located inside $P_{x,y}$ (the sequence cannot ``jump over $P_{x,y}$'', as then the distance would be too large).
		Moreover, if we truncate this sequence at $x'$, we get a~witness that $x'\in Z_j$, for some $j\in\{1,\dots,k\}$.
		Simultaneously, still from the definition of $Z_i$, we obtain $\stab(z,i)<\stab(x',j)$.
		Denoting $\xi=\max\{\stab(x',j)-1\mid j\in\{1,\dots,k\},x'\in P_{x,y}\cap Z_j\}$ we thus have $\stab(z,i)\leq\xi$ for all $i\in\{1,\dots,k\}$ and $z\in Z_i\restr_y$
		(the ``$-1$'' operation is well\=/defined because $\stab(x',j)$ is a~successor ordinal).
		By assumption we have $\stab(x',j)\leq\gamma$ for $x'\in P_{x,y}\cap Z_j\subseteq Z_j\restr_x$, that is $\xi<\gamma$.
		The induction hypothesis then implies that $\rank(U\restr_y)\leq D\cdot(\xi+1)\leq D\cdot\gamma$.

		Thus, we have $\rank(U\restr_y)\leq D\cdot\gamma$ for every node $y$ such that $\dist(x,y)=D$.
		Between $x$ and those nodes $y$ we additionally have $D$ layers of nodes (some of them being in $U\restr_x$, some not).
		It can be easily seen that such $D$ layers may increase the rank at most by $D$, so $\rank(U\restr_x)\leq D\cdot(\gamma+1)$, as claimed.
	\end{proofof}
	
	We now move towards the proof of the implication ``$\Leftarrow$'' of \cref{lem:mu-tree2}.
	Recall the assumption: for some limit ordinal $\eta$ and some $I\subseteq\{1,\dots,k\}$
	there exists a~tuple $\vec{Z}$ of sets $Z_1,\dots,Z_k\subseteq\dom(\tau)$ such that $\vec{F}(\vec{Z})\supseteq\vec{Z}$, and $\rank(Z_1\cup\dots\cup Z_k)\leq\eta$, and $r\in Z_i$ for all $i\in I$.
	Again, denote $U=Z_1\cup\dots\cup Z_k$.
	The thesis, saying that $r\in\pi_i(\mu^\eta\vec{X}.\,\vec{F}(\vec{X}))$ for all $i\in I$ is obtained from the following claim by taking $\gamma=\eta$ and $x=r$.

	\begin{clm}
		For every index $i$, ordinal $\gamma$, and node $x$ such that $\rank(U\restr_x)\leq\gamma$ it holds $Z_i\restr_x\subseteq\pi_i(\mu^\gamma\vec{X}.\,\vec{F}(\vec{X}))$.
	\end{clm}

	\begin{proof}
		The proof is by induction on $\gamma$.
		Consider first the special case when $x\in Z_i\subseteq U$.
		Then, by definition, $\rank(U\restr_x)$ is the number of a~step in which the derivative removes the node $x$; it is thus necessarily a~successor ordinal $\xi+1\leq\gamma$.
		Moreover, for every successor $y$ of $x$ we then have $\rank(U\restr_y)\leq\xi$,
		so the induction hypothesis implies that $Z_j\restr_y\subseteq\pi_j(\mu^{\xi}\vec{X}.\,\vec{F}(\vec{X}))$ for every $j\in\{1,\dots,k\}$.
		Taking the union over all successors, we obtain $Z_j\restr_x\setminus\{x\}\subseteq\pi_j(\mu^{\xi}\vec{X}.\,\vec{F}(\vec{X}))$.
		Denote $\vec{Z}'=(Z_1\restr_x\setminus\{x\},\dots,Z_k\restr_x\setminus\{x\})$.
		Now, note that the value of $F_i(\vec{X})$ in nodes reachable from $x$ depends on the valuation of the variables $X_1,\dots,X_k$ only in nodes reachable from $x$.
		Moreover, we assumed that those variables occur in $F_i(\vec{X})$ only in the scope of modal operators, meaning that they are not accessed in the node $x$ itself.
		In other words, $F_i(\vec{Z})$ agrees with $F_i(\vec{Z}')$ on all nodes reachable from $x$.
		Thus, from $Z_i\subseteq F_i(\vec{Z})$ (and then from monotonicity of $F_i$) we can deduce that
		\begin{align*}
			Z_i\restr_x\subseteq F_i(\vec{Z}')\subseteq F_i(\mu^{\xi}\vec{X}.\,\vec{F}(\vec{X}))=\pi_i(\mu^{\xi+1}\vec{X}.\,\vec{F}(\vec{X}))\subseteq\pi_i(\mu^\gamma\vec{X}.\,\vec{F}(\vec{X})).
		\end{align*}
		
		The general case, when $x\not\in Z_i$ can be easily reduced to the previous one,
		because $Z_i\restr_x$ is the union of $Z_i\restr_y$ over all nodes $y\in Z_i$ reachable from $x$;
		moreover $\rank(U\restr_y)\leq\rank(U\restr_x)\leq\gamma$ for such nodes $y$.
	\end{proof}

	This finishes the proof of \cref{lem:mu-tree2}.
\end{proof}

Composing equivalences from \cref{lem:mu-tree1,lem:mu-tree2} we obtain the following \lcnamecref{cor:mu-tree}.

\begin{cor}\label{cor:mu-tree}
	Let $\vec{F}(\vec{X})=(F_1(\vec{X}),\dots,F_k(\vec{X}))$ with $\vec{X}=(X_1,\dots,X_k)$ be a~tuple of formulae as in \cref{thm:closure-mu}, with all occurrences of $X_1,\dots,X_k$ being in the scope of a~modal operator.
	The following two conditions are equivalent for every countable limit ordinal $\eta$:
	\begin{itemize}
	\item	the closure ordinal of $\mu \vec{X}.\,\vec{F}(\vec{X})$ is bounded by $\eta$;
	\item	for every model $\tau$ that is a~countable tree with root $r$ there exists a~tuple $\vec{Z}$ of well\=/founded sets $Z_1,\dots,Z_k\subseteq\dom(\tau)$
		such that $\vec{F}(\vec{Z})\supseteq\vec{Z}$, and $\rank(Z_1\cup\dots\cup Z_k)\leq\eta$, and $r\in Z_i$ for all $i\in\{1,\dots,k\}$ such that $r\in\pi_i(\mu \vec{X}.\,\vec{F}(\vec{X}))$.
	\end{itemize}
\end{cor}
\noindent 
A~countable tree $\tau$, occurring in the \lcnamecref{cor:mu-tree} above, can be seen as a~function $\tau\from X\to A$ from a~prefix\=/closed subset $X\subseteq\N^\ast$ to a~finite alphabet $A$.
Now, recall a~natural encoding $(n_1,n_2,\ldots,n_k)\mapsto \dR^{n_1}\dL\dR^{n_2}\dL\cdots\dR^{n_k}\dL$ of $\N^\ast$ into $\{\dL,\dR\}^\ast$.
This encoding preserves the prefix order on $\N^\ast$ and moreover preserves ranks of well\=/founded sets.
Take the relation $\Gamma_{\!\!\vec{F}}$ that contains $(t,x)\in\trees_A\times\trees_{\{0,1\}}$ if:
\begin{itemize}
\item	$t$ encodes a~model $\tau$ with root $r$,
\item	$x$ encodes a~well\=/founded set $U\subseteq\dom(\tau)$,
\item	there exists a~tuple $\vec{Z}=(Z_1,\dots,Z_k)$ such that $Z_1\cup\dots\cup Z_k=U$, and $\vec{F}(\vec{Z})\supseteq\vec{Z}$,
	and $r\in Z_i$ for all $i\in\{1,\dots,k\}$ such that $r\in\pi_i(\mu \vec{X}.\,\vec{F}(\vec{X}))$.
\end{itemize}
The $\mu$\=/calculus formulae $\vec{F}(\vec{Z})$ and $\mu\vec{X}.\,\vec{F}(\vec{X})$ can be rewritten into MSO~\cite{julian-igor-handook}
and then modified to read the above encoding of $\tau$ in the binary tree, instead of $\tau$ itself.
It follows that the relation $\Gamma_{\!\!\vec{F}}$ is MSO\=/definable.

Observe that the second item of \cref{cor:mu-tree} can be rephrased by saying that the closure ordinal of $\Gamma_{\!\!\vec{F}}$ is bounded by $\eta$.
Applying \cref{thm:main} to $\Gamma_{\!\!\vec{F}}$, and then \cref{cor:mu-tree}, we have one of two possibilities:
If the closure ordinal of $\Gamma_{\!\!\vec{F}}$ is smaller than $\omega^2$, then it is bounded by $\omega\cdot N$ for some $N\in\N$;
then also the closure ordinal of $\mu\vec{X}.\,\vec{F}(\vec{X})$ is bounded by $\omega\cdot N<\omega^2$.
Otherwise, the closure ordinal of $\Gamma_{\!\!\vec{F}}$ is $\omega_1$, so it is not bounded by any countable limit ordinal;
then also the closure ordinal of $\mu\vec{X}.\,\vec{F}(\vec{X})$ is not bounded by any countable limit ordinal, hence it is at least $\omega_1$.
This concludes the proof of \cref{thm:closure-mu}.

\begin{rem}
	One may ask if \cref{thm:closure-mu} is merely a~consequence or it is in some sense equivalent to our main result \cref{thm:main}.
	To the best of our understanding, \cref{thm:closure-mu} does not transfer back to the general realm of MSO\=/definable relations, as in \cref{thm:main}.
	One of the reasons is that the iterations of fixed points are required to proceed in a~monotone fashion, driven by the internal formulae $\vec{F}$;
	while in full MSO one can express arbitrary correspondence between the parameters $\vec{Y}$ and a~well\=/founded witness $X$.
\end{rem}

\section{Conclusions}
\label{sec:conclusions}

This work contributes to the study of expressive power of the MSO theory of the binary tree.
We investigate to what extent this theory can express properties of well\=/founded trees, and in particular distinguish between their ordinal ranks.
We observe that the ability to express properties of ranks explicitly is practically limited to statements of the form:
all trees $X$ satisfying $\varphi (X)$ have $\rank (X) < N$, for a~fixed $N \in \N$ (cf.\ \cref{cor:not-expressible} above).
However, the implicit expressive power of MSO logic goes much higher.
In particular, our main result (\cref{thm:main}) allows us to decide whether the property
\[
  \exists X. \, \varphi (\vec{Y}, X) \; \wedge \;
  X \mbox{ is well\=/founded with } \rank (X) < \omega^2,
\]
is generally true (for all $\vec{Y}$), although the property itself is not expressible in MSO.

There are, however, a~number of questions that remain to be answered.
As ordinals smaller than~$\omega^2$ can be effectively represented, we would like to have an~effective procedure that, given a~formula $\varphi$, computes the exact bound,
that is, (a~representation of) $\rank(\varphi)$.
Even more elementarily, given an~MSO\=/definable set $L$ of
well\=/founded trees, we would like to compute the supremum of ranks of trees in $L$. Although these problems are open for $\rank$, the present results show that they are computable for $\rankR$, indicating that it might be a~more robust rank in the context of MSO.

A~more far\=/reaching direction is to relate the techniques of the present paper to the open problem of computing the Mostowski index, mentioned in Introduction.
The parity condition itself imposes well\=/foundedness restriction on the occurrences of each odd label $m$ in the fragments of the tree where this label is the highest.
Colcombet and Löding~\cite{loding_index_to_bounds} have approached the index problem (still unsolved) by reducing it to the boundedness problem for distance automata
(see also Idir and Lehtinen~\cite{lehtinen_mostowski} for a~more direct version of this reduction).
One may consider an~alternative approach towards the index problem by studying the ordinal ranks which arise from the well\=/foundedness restriction of the parity condition.

\section*{Acknowledgement}\noindent
The authors would like to thank Marek Czarnecki for preliminary discussions on related subjects.
They would also like to thank the anonymous reviewers of both the conference and journal versions of the paper for their valuable comments and suggestions.

\bibliographystyle{alphaurl}
\bibliography{mskrzypczak}

@inproceedings{bojanczyk_wmso_u_p,
  author       = {Miko{\l}aj Bojańczyk},
  title        = {Weak {MSO+U} with Path Quantifiers over Infinite Trees},
  booktitle    = {{ICALP} {(2)}},
  doi          = {10.1007/978-3-662-43951-7_4},
  series       = {Lecture Notes in Computer Science},
  volume       = {8573},
  pages        = {38--49},
  publisher    = {Springer},
  year         = {2014}
}

@inproceedings{ong_model_checking_higher_order,
  author       = {C.{-}H. Luke Ong},
  title        = {On Model-Checking Trees Generated by Higher-Order Recursion Schemes},
  booktitle    = {LICS},
  pages        = {81--90},
  publisher    = {{IEEE} Computer Society},
  year         = {2006},
  doi          = {10.1109/LICS.2006.38},
}

@inproceedings{Hausmann2022,
author="Hausmann, Daniel
and Piterman, Nir",
title="A Survey on Satisfiability Checking for the $\mu$-Calculus Through Tree Automata",
bookTitle="Principles of Systems Design",
year="2022",
publisher="Springer",
pages="228--251",
isbn="978-3-031-22337-2",
doi="10.1007/978-3-031-22337-2_11",
}

@book{kechris_descriptive,
  author      = {Alexander Kechris},
  title       = {Classical descriptive set theory},
  publisher   = {Springer-Verlag},
  address     = {New York},
  year        = {1995}
}

@book{niwinski_rudiments,
  author      = {Andr{\'{e}} Arnold and
                 Damian Niwi{\'{n}}ski},
  title       = {Rudiments of mu-calculus},
  publisher   = {Elsevier},
  series      = {Studies in Logic and the Foundations of Mathematics},
  year        = {2001}
}

@inproceedings{Bekic1984,
  author       = {Hans Beki{\'{c}}},
  title        = {Definable Operation in General Algebras, and the Theory of Automata
                  and Flowcharts},
  booktitle    = {Programming Languages and Their Definition},
  series       = {Lecture Notes in Computer Science},
  volume       = {177},
  pages        = {30--55},
  publisher    = {Springer},
  year         = {1984},
  doi          = {10.1007/BFB0048939},
}

@book{Demri2016,
  author    = {St{\'{e}}phane Demri and
               Valentin Goranko and
               Martin Lange},
  title     = {Temporal Logics in Computer Science: Finite-State Systems},
  series    = {Cambridge Tracts in Theoretical Computer Science},
  publisher = {Cambridge University Press},
  year      = {2016}
}

@book{mskrzypczak_lncs,
  author      = {Micha{\l} Skrzypczak},
  title       = {Descriptive Set Theoretic Methods in Automata Theory -- Decidability and Topological Complexity},
  series      = {Lecture Notes in Computer Science},
  volume      = {9802},
  publisher   = {Springer},
  year        = {2016},
  isbn        = {978-3-662-52946-1},
  doi         = {10.1007/978-3-662-52947-8},
}

@article{santocanale-closure,
  author       = {Maria J. Gouveia and
                  Luigi Santocanale},
  title        = {{\(\aleph\)}\({}_{\mbox{1}}\) and the modal {\(\mu\)}-calculus},
  journal      = {Log. Methods Comput. Sci.},
  volume       = {15},
  number       = {4},
  year         = {2019},
  doi          = {10.23638/LMCS-15(4:1)2019}
}

@article{barany_expressing_trees,
  author      = {Vince B{\'{a}}r{\'{a}}ny and
                 {\L}ukasz Kaiser and
                 Alexander M. Rabinovich},
  title       = {Expressing Cardinality Quantifiers in Monadic Second-Order Logic over Trees},
  journal     = {Fundam. Inform.},
  volume      = {100},
  number      = {1-4},
  pages       = {1--17},
  year        = {2010},
  doi         = {10.3233/FI-2010-260}
}

@article{buchi_synthesis,
  author      = {Julius R. B{\"{u}}chi and
                 Lawrence H. Landweber},
  title       = {Solving Sequential Conditions by Finite-State Strategies},
  journal     = {Transactions of the American Mathematical Society},
  pages       = {295--311},
  publisher   = {American Mathematical Society},
  volume      = {138},
  year        = {1969}
}

@article{esparza-acceleration,
  author       = {Javier Esparza and
                  Stefan Kiefer and
                  Michael Luttenberger},
  title        = {Derivation tree analysis for accelerated fixed-point computation},
  journal      = {Theor. Comput. Sci.},
  volume       = {412},
  number       = {28},
  pages        = {3226--3241},
  year         = {2011},
  doi          = {10.1016/J.TCS.2011.03.020},
}

@article{finkel-todo-auto,
  author       = {Olivier Finkel and
                  Stevo Todorcevic},
  title        = {Automatic Ordinals},
  journal      = {Int. J. Unconv. Comput.},
  volume       = {9},
  number       = {1-2},
  pages        = {61--70},
  year         = {2013},
  url          = {http://www.oldcitypublishing.com/journals/ijuc-home/ijuc-issue-contents/ijuc-volume-9-number-1-2-2013/ijuc-9-1-2-p-61-70/},
}

@article{delhomme,
  author       = {Christian Delhomm\'e},
  title        = {Automaticit\'e des ordinaux et des graphes homog\`enes},
  journal      = {Comptes Rendus de L’Acad\'emie des Sciences, Math\'ematiques},
  volume       = {339},
  number       = {1},
  pages        = {5--10},
  year         = {2004}
}

@article{kozen-mu-calc,
  author       = {Dexter Kozen},
  title        = {Results on the Propositional mu-Calculus},
  journal      = {Theor. Comput. Sci.},
  volume       = {27},
  pages        = {333--354},
  year         = {1983},
  doi          = {10.1016/0304-3975(82)90125-6},
  bibsource    = {dblp computer science bibliography, https://dblp.org}
}

@article{niwinski_gap,
  author      = {Damian Niwi{\'{n}}ski and
                 Igor Walukiewicz},
  title       = {A gap property of deterministic tree languages},
  journal     = {Theor. Comput. Sci.},
  volume      = {1},
  number      = {303},
  year        = {2003},
  pages       = {215--231},
  doi         = {10.1016/S0304-3975(02)00452-8},
}

@article{rabin_s2s,
  author      = {Michael O. Rabin},
  title       = {Decidability of second-order theories and automata on infinite trees},
  journal     = {Transactions of the American Mathematical Society},
  volume      = {141},
  year        = {1969},
  pages       = {1--35}
}

@inproceedings{afshari_closure,
  author      = {Bahareh Afshari and
                 Graham E. Leigh},
  title       = {On closure ordinals for the modal mu-calculus},
  booktitle    = {{CSL}},
  series       = {LIPIcs},
  volume       = {23},
  pages        = {30--44},
  publisher    = {Schloss Dagstuhl - Leibniz-Zentrum f{\"{u}}r Informatik},
  year         = {2013},
  doi          = {10.4230/LIPICS.CSL.2013.30},
}

@inproceedings{bradfield_simplifying,
  author      = {Julian C. Bradfield},
  title       = {Simplifying the modal mu-calculus alternation hierarchy},
  series      = {Lecture Notes in Computer Science},
  volume      = {1373},
  booktitle   = {STACS},
  publisher   = {Springer}, 
  year        = {1998},
  pages       = {39--49},
  doi         = {10.1007/BFB0028547}
}

@InProceedings{bruse_closure,
  author =	{Bruse, Florian and S\"{a}lzer, Marco and Lange, Martin},
  title =	{Finite Convergence of $\mu$-Calculus Fixpoints on Genuinely Infinite Structures},
  booktitle =	{{MFCS}},
  pages =	{24:1--24:19},
  series =	{LIPIcs},
  ISBN =	{978-3-95977-201-3},
  ISSN =	{1868-8969},
  year =	{2021},
  volume =	{202},
  publisher =	{Schloss Dagstuhl - Leibniz-Zentrum f{\"u}r Informatik},
  doi =		{10.4230/LIPIcs.MFCS.2021.24}
}

@inproceedings{clemente_separability,
  author       = {Lorenzo Clemente and
                  Micha{\l} Skrzypczak},
  title        = {Deterministic and Game Separability for Regular Languages of Infinite Trees},
  booktitle    = {ICALP},
  series       = {LIPIcs},
  volume       = {198},
  pages        = {126:1--126:15},
  publisher    = {Schloss Dagstuhl - Leibniz-Zentrum f{\"{u}}r Informatik},
  doi          = {10.4230/LIPICS.ICALP.2021.126},
  year         = {2021}
}

@inproceedings{colcombet_weak,
  author      = {Thomas Colcombet and
                 Denis Kuperberg and
                 Christof L{\"{o}}ding and
                 Michael {Vanden Boom}},
  title       = {Deciding the weak definability of {B{\"{u}}chi} definable tree languages},
  booktitle   = {CSL},
  year        = {2013},
  pages       = {215--230},
  series      = {LIPIcs},
  volume      = {23},
  publisher   = {Schloss Dagstuhl - Leibniz-Zentrum f{\"{u}}r Informatik},
  doi         = {10.4230/LIPICS.CSL.2013.215},

}

@inproceedings{bradfield_transfinite,
  author      = {Julian C. Bradfield and
                 Jacques Duparc and
                 Sandra Quickert},
  title       = {Transfinite Extension of the mu-Calculus},
  booktitle   = {CSL},
  year        = {2005},
  series      = {Lecture Notes in Computer Science},
  volume      = {3634},
  publisher   = {Springer},
  pages       = {384--396},
  doi         = {10.1007/11538363_27},
}

@inproceedings{fontaine_continuous,
  author      = {Ga{\"e}lle Fontaine},
  title       = {Continuous Fragment of the mu-Calculus},
  booktitle   = {CSL},
  year        = {2008},
  pages       = {139--153},
  series      = {Lecture Notes in Computer Science},
  volume      = {5213},
  publisher   = {Springer},
  doi         = {10.1007/978-3-540-87531-4_12},
}

@inproceedings{loding_index_to_bounds,
  author      = {Thomas Colcombet and
                 Christof L{\"{o}}ding},
  title       = {The Non-deterministic {Mostowski} Hierarchy and Distance-Parity Automata},
  booktitle   = {ICALP (2)},
  year        = {2008},
  pages       = {398--409},
  series      = {Lecture Notes in Computer Science},
  volume      = {5126},
  publisher   = {Springer},
  doi         = {10.1007/978-3-540-70583-3_33}
}

@inproceedings{murlak_game_auto,
  author      = {Alessandro Facchini and
                 Filip Murlak and
                 Micha{\l} Skrzypczak},
  title       = {{Rabin-Mostowski} Index Problem: A Step beyond Deterministic Automata},
  publisher   = {{IEEE} Computer Society},
  booktitle   = {LICS},
  year        = {2013},
  pages       = {499--508},
  doi         = {10.1109/LICS.2013.56},
}

@inproceedings{niwinski_cardinality,
  author      = {Damian Niwi{\'{n}}ski},
  title       = {On the Cardinality of Sets of Infinite Trees Recognizable by Finite Automata},
  booktitle   = {MFCS},
  pages       = {367--376},
  year        = {1991},
  series      = {Lecture Notes in Computer Science},
  volume      = {520},
  publisher   = {Springer},
  doi         = {10.1007/3-540-54345-7_80},
}

@inproceedings{jutla_determinacy,
  author      = {Allen Emerson and
                 Charanjit Jutla},
  title       = {Tree Automata, mu-Calculus and Determinacy},
  booktitle   = {FOCS},
  pages       = {368--377},
  publisher   = {{IEEE} Computer Society},
  year        = {1991},
  doi         = {10.1109/SFCS.1991.185392},
}

@inproceedings{dichotomy-stacs,
  author       = {Damian Niwiński and
                  Paweł Parys and
                  Michał Skrzypczak},
  title        = {A Dichotomy Theorem for Ordinal Ranks in {MSO}},
  booktitle    = {{STACS}},
  pages        = {69:1--69:18},
  year         = {2025},
  doi          = {10.4230/LIPICS.STACS.2025.69},
  series       = {LIPIcs},
  volume       = {327},
  publisher    = {Schloss Dagstuhl - Leibniz-Zentrum f{\"{u}}r Informatik},
}

@inproceedings{walukiewicz_buchi,
  author      = {Micha{\l} Skrzypczak and
                 Igor Walukiewicz},
  title       = {Deciding the Topological Complexity of {B}{\"{u}}chi Languages},
  booktitle    = {{ICALP}},
  series       = {LIPIcs},
  volume       = {55},
  pages        = {99:1--99:13},
  publisher    = {Schloss Dagstuhl - Leibniz-Zentrum f{\"{u}}r Informatik},
  year         = {2016},
  doi          = {10.4230/LIPICS.ICALP.2016.99}
}

@inproceedings{thomas_languages,
  author      = {Wolfgang Thomas},
  title       = {Languages, Automata, and Logic},
  booktitle   = {Handbook of Formal Languages {(3)}},
  year        = {1997},
  pages       = {389--455},
  publisher   = {Springer},
  doi         = {10.1007/978-3-642-59126-6_7},
}

@incollection{julian-igor-handook,
  author       = {Julian C. Bradfield and
                  Igor Walukiewicz},
  title        = {The mu-calculus and Model Checking},
  booktitle    = {Handbook of Model Checking},
  pages        = {871--919},
  year         = {2018},
  publisher    = {Springer},
  doi          = {10.1007/978-3-319-10575-8_26},
}

@techreport{mostowski_parity_games,
  author      = {Andrzej W. Mostowski},
  title       = {Games with forbidden positions},
  institution = {University of Gda{\'{n}}sk},
  year        = {1991}
}

@inproceedings{czarnecki_closure_fics,
  author       = {Marek Czarnecki},
  title        = {How Fast Can the Fixpoints in Modal mu-Calculus Be Reached?},
  booktitle    = {{FICS}},
  pages        = {35--39},
  publisher    = {Laboratoire d'Informatique Fondamentale de Marseille},
  year         = {2010},
  url          = {https://hal.archives-ouvertes.fr/hal-00512377/document\#page=36}
}

@inproceedings{afshari_limit_fics,
  author       = {Bahareh Afshari and Giacomo Barlucchi and Graham E. Leigh},
  title        = {The Limit of Recursion in State-based Systems},
  booktitle    = {{FICS}},
  year         = {2025},
  series    = {Electronic Proceedings in Theoretical Computer Science},
  volume    = {435},
  publisher = {Open Publishing Association},
  pages     = {1-12},
  doi       = {10.4204/EPTCS.435.1},
}

@article{fijalkow_games,
  author       = {Nathana{\"{e}}l Fijalkow and
                  Nathalie Bertrand and
                  Patricia Bouyer{-}Decitre and
                  Romain Brenguier and
                  Arnaud Carayol and
                  John Fearnley and
                  Hugo Gimbert and
                  Florian Horn and
                  Rasmus Ibsen{-}Jensen and
                  Nicolas Markey and
                  Benjamin Monmege and
                  Petr Novotn{\'{y}} and
                  Mickael Randour and
                  Ocan Sankur and
                  Sylvain Schmitz and
                  Olivier Serre and
                  Mateusz Skomra},
  title        = {Games on Graphs},
  journal      = {CoRR},
  volume       = {abs/2305.10546},
  year         = {2023},
  doi          = {10.48550/ARXIV.2305.10546},
  bibsource    = {dblp computer science bibliography, https://dblp.org}
}

@phdthesis{schmitz_hab,
  author       = {Sylvain Schmitz},
  title        = {Algorithmic Complexity of Well-Quasi-Orders. (Complexit{\'{e}}
                  algorithmique des beaux pr{\'{e}}-ordres)},
  school       = {Université Paris Diderot (Paris 7)},
  type         = {Habilitation thesis},
  year         = {2017},
  publisher    = {Université Paris Diderot},
  url          = {https://tel.archives-ouvertes.fr/tel-01663266}
}

@article{lehtinen_mostowski,
       title={Mostowski Index via extended register games},
       author={Olivier Idir and Karoliina Lehtinen},
  journal      = {CoRR},
  volume       = {abs/2412.16793},
  year         = {2024},
  doi          = {10.48550/ARXIV.2412.16793},
}

@misc{lange-convergence,
      title={Finite Convergence of the Modal Mu-Calculus on Almost-Periodic Words}, 
      author={Fabian Lehr and Florian Bruse},
      year={2026},
      eprint={2607.08181},
      archivePrefix={arXiv},
      primaryClass={cs.LO},
      url={https://arxiv.org/abs/2607.08181}, 
}
\end{document}